\documentclass[conference]{IEEEtran}

\ifCLASSINFOpdf
\else
\fi
\usepackage[hyphens]{url}
\usepackage{amsmath}
\usepackage{amsfonts}
\usepackage{amssymb}
\usepackage{bm}

\usepackage{cite}
\usepackage{color}
\usepackage[pdftex]{graphicx}
\usepackage{threeparttable}
\usepackage{subfig}
\usepackage{multirow}
\usepackage{float}

\usepackage[]{fancyhdr} %

\IEEEoverridecommandlockouts
\begin{document}

%
\title{Optimizing Transmission Infrastructure Investments to Support Line De-energization for\\ Mitigating Wildfire Ignition Risk}

\author{\IEEEauthorblockN{Alyssa Kody}
\IEEEauthorblockA{Energy Systems Division\\Argonne National Laboratory\\
Lemont, IL, USA\\
akody@anl.gov}\vspace*{-2em}
\and
\IEEEauthorblockN{Ryan Piansky, Daniel K. Molzahn}
\IEEEauthorblockA{School of Electrical and Computer Engineering\\Georgia Institute of Technology\\
Atlanta, GA, USA\\
\{rpiansky3,molzahn\}@gatech.edu}\vspace*{-2em}\thanks{This work is partially supported by the U.S. Department of Energy, Office of Electricity and Laboratory Directed Research and Development (LDRD) funding from Argonne National Laboratory, provided by the Director, Office of Science, of the U.S. Department of Energy under Contract No. DE-AC02-06CH11357. This work is also partially supported by the U.S. National Science Foundation Energy, Power, Control and Networks (EPCN) under award \#2145564.}
}

\lhead{\scriptsize{ACCEPTED FOR PRESENTATION IN 11TH BULK POWER SYSTEMS DYNAMICS AND CONTROL SYMPOSIUM, JULY 25-30, 2022, BANFF, CANADA}}


\maketitle
\thispagestyle{fancy}
\pagestyle{fancy}


\begin{abstract}
Wildfires pose a growing risk to public safety in regions like the western United States, and, historically, electric power systems have ignited some of the most destructive wildfires. To reduce wildfire ignition risks, power system operators preemptively de-energize high-risk power lines during extreme wildfire conditions as part of ``Public Safety Power Shutoff’’ (PSPS) events. While capable of substantially reducing acute wildfire risks, PSPS events can also result in significant amounts of load shedding as the partially de-energized system may not be able to supply all customer demands. In this work, we investigate the extent to which infrastructure investments can support system operations during PSPS events by enabling reduced load shedding and wildfire ignition risk. We consider the installation of grid-scale batteries, solar PV, and line hardening or maintenance measures (e.g., undergrounding or increased vegetation management). Optimally selecting the locations, types, and sizes of these infrastructure investments requires considering the line de-energizations associated with PSPS events. Accordingly, this paper proposes a multi-period optimization formulation that locates and sizes infrastructure investments while simultaneously choosing line de-energizations to minimize wildfire ignition risk and load shedding. The proposed formulation is evaluated using two geolocated test cases along with realistic infrastructure investment parameters and actual wildfire risk data from the 
United States Geological Survey. We evaluate the performance of investment choices by simulating de-energization decisions for the entire 2021 wildfire season with optimized infrastructure placements. With investment decisions varying significantly for different test cases, budgets, and operator priorities, the numerical results demonstrate the proposed formulation’s value in tailoring investment choices to different settings.

\end{abstract}



%
\IEEEpeerreviewmaketitle

\section{Introduction} \label{sec:intro}
Climate change is increasing the prevalence of wildfire-prone conditions, leading to more severe and frequent wildfires~\cite{noaa,goss2020climate}. While most wildfires are not started by electric power infrastructure, wildfires ignited by power lines tend to be more destructive than those from other sources~\cite{keeley2019twenty}. For instance, less than 10\% of reported wildfire ignitions in California are due to power lines, but these account for about half of the most destructive fires~\cite{psps_cpuc}, including the 2018 Camp Fire that is classified as the ``most destructive fire in California history'' with over 100,000 burnt acres, 85 deaths, and over ten billion dollars in damage~\cite{muhs2020wildfire, CALFIRE_campfire}. 
Accordingly, engineers must operate and design power systems in a manner that mitigates the risk of wildfire ignitions~\cite{muhs2020wildfire,arab2021,vazquez2022}.

To address imminent wildfire conditions, system operators use so-called ``Public Safety Power Shutoffs'' (PSPS) that temporarily de-energize power lines located in wildfire-prone regions during severe wildfire conditions~\cite{psps_cpuc}. De-energized lines cannot ignite wildfires, so PSPS strategies are effective at quickly reducing acute wildfire ignition risks. However, risk reduction from PSPS events comes at the cost of load shedding if the partially de-energized system cannot supply all load demands. 
A recent cost/benefit analysis in~\cite{lafollette_cba} regarding various strategies for mitigating wildfire ignition risks indicates that PSPS events are a cost-effective mechanism for reducing acute wildfire risks. Since they are likely to continue being employed during severe wildfire-prone conditions, 
PSPS events deserve further research to achieve system operators' goals of making PSPS events ``smaller in scope, shorter in duration, and smarter in performance''~\cite{PGE_WildfirePlan}.

To optimally balance reductions in wildfire ignition risks and load shedding, Rhodes, Ntaimo, and Roald have recently formulated and solved optimal transmission switching (OTS) problems that determine which lines to de-energize~\cite{rhodes2020balancing}. Optimizing PSPS events via this OTS approach can significantly reduce both wildfire ignition risks and load shedding when compared to the alternative of de-energizing all lines above a specified risk threshold.

The key difference between the OTS formulation in~\cite{rhodes2020balancing} and previous OTS formulations (see, e.g.,~\cite{fisher2008optimal,hedman2011survey}) is that the lines are switched off solely for the sake of de-energizing them to reduce wildfire ignition risk, not to achieve some other objective like reducing operating costs~\cite{fisher2008optimal,fuller2012,johnson2021} or improving reliability~\cite{lyon2016,li2017}. Prior OTS applications achieved these other objectives by focusing on the ability of transmission switching to ameliorate network congestion. In contrast, the sets of lines de-energized to mitigate wildfire ignition risks are not strongly related to network congestion effects. Moreover, in contrast to prior OTS applications, OTS problems in a wildfire context must consider the possibility of significant load shedding resulting from de-energization of non-negligible portions of the system.

Over longer time scales, infrastructure investments can reduce the amount of load shedding needed to achieve desired risk reductions during PSPS events. For instance, utilities can:
\begin{itemize}
    \item Add batteries that supply loads during PSPS events to reduce the amount of load shedding,
    \item Install distributed energy resources such as solar photovoltaic generators to provide local supplies of power,
    \item Harden power lines via undergrounding, installing covered conductors, and performing intensive vegetation management in order to reduce wildfire ignition risk without the need for de-energization.
\end{itemize}

The severe impacts of wildfires justify 
substantial infrastructure investments. For instance, the Infrastructure Investment and Jobs Act recently passed in the United States allocates \$5 billion towards resiliency measures like undergrounding power lines to prevent wildfire ignitions and installing microgrids to reduce the impacts of PSPS events~\cite{infrastructure_bill}. Likewise, in April 2021, the state of California funded \$536 million for wildfire resilience projects~\cite{ca_infrastructure_bill}. Moreover, the California utility PG\&E plans to underground 10,000 miles of power lines to prevent wildfire ignitions, which will require a substantial increase in the utility's current rate of 70 miles of newly undergrounded lines per year~\cite{pge_undergrounding_initiative,pge_undergrounding_initiative_nyt}.

The scale of these investments motivates the development of algorithms for optimally siting and sizing new power system infrastructure in a wildfire context. There is an extensive literature on siting and sizing various power system components with the aim of reducing generation costs and providing ancillary services~\cite{dj2014hicss,paliwal2014,hrvoje2015,wogrin2015,fiorini2017,baker2017,wullner2021review}, improving reliability~\cite{paliwal2014}, deferring capital investments~\cite{paliwal2014,wullner2021review}, etc.
However, the wildfire setting presents a key challenge that differs from this prior literature, namely, that the \emph{infrastructure investments will be operated to support future PSPS events}. Thus, choices for the locations and sizes of the infrastructure investments should be cognizant of the underlying line de-energizations associated with PSPS events. 

To the best of our knowledge, none of the existing wildfire risk mitigation literature (see~\cite{vazquez2022} for a recent survey) proposes infrastructure investment algorithms that consider OTS-based line de-energization using a power flow model as in~\cite{rhodes2020balancing}. In other words, none of the existing literature discusses how to optimally plan infrastructure investments in order to support system operations during PSPS events. A cost/benefit analysis by Williamson indicates that solar PV generation could be effective at reducing load shedding during PSPS events in Australia~\cite{williamson2015}, Haces-Fernandez studies the suitability of wind generators to locally supply power during wildfire-prone conditions~\cite{haces2020}, and Taylor and Roald consider line undergrounding investments in the context of various wildfire risk metrics~\cite{taylor2021wildfire}. While providing many valuable insights, the formulations in these papers do not incorporate a power flow model and may therefore miss important spatial interactions and network constraints inherent to power systems. Other papers focus on methods for operating power systems during wildfire-prone conditions. For instance, both Nazemi et al. and Tandon, Grijalva, and Molzahn study the impact of dynamic line ratings to increase operational flexibility~\cite{nazemi2021,tandon_grijalva_molzahn-peci2021}, Hong et al. propose data-driven techniques for minimizing load shedding after switching off high-risk lines while considering the possibility of cascading failures~\cite{hong2022hicss}, Zhou et al. use data-mining techniques to assess and mitigate wildfire ignition risks~\cite{zhou2019}, Haseltine and Roald analyze how recloser operation affects both wildfire risks and system reliability~\cite{haseltine2021}, and Kadir et al. describe a reinforcement learning approach to line de-energization and other operational decisions~\cite{kadir2021}. However, none of these papers incorporate an infrastructure investment model. Many other papers propose methods for enhancing power system resilience to extreme events such as wildfires, but do not explicitly consider a line de-energization model based on wildfire ignition risks; see~\cite{panteli2015,jufri2019,mahzarnia2020,bhusal2020} for recent surveys of the power system resilience literature.

Accordingly, this paper proposes optimization formulations that augment OTS problems which minimize wildfire ignition risks with models for optimally siting and sizing various infrastructure investments. We specifically consider investments in batteries, solar photovoltaic (PV) generators, and infrastructure hardening via undergrounded lines, covered conductors, and intensive vegetation management. We first propose a multi-period extension of the OTS problem presented in~\cite{rhodes2020balancing} for mitigating the risk of wildfire ignitions. The time periods in this problem are coupled by both the selection of a fixed network topology across all periods and the batteries' state-of-charge dynamics. We then extend this multi-period OTS problem to an infrastructure investment formulation that incorporates models of batteries, solar PV generators, and infrastructure hardening, using discrete variables to represent the presence of these investments at each location. 
Finally, we numerically demonstrate the proposed infrastructure investment formulation using realistic test cases with actual wildfire risk and infrastructure investment data.

As in much of the other existing literature on mitigating wildfire ignition risks (e.g.,~\cite{zhou2019,rhodes2020balancing,tandon_grijalva_molzahn-peci2021,taylor2021wildfire,hong2022hicss,kadir2021}), we focus on transmission systems. Transmission systems have ignited major fires, such as the 2018 Camp Fire in California~\cite{muhs2020wildfire, CALFIRE_campfire}, and are thus deserving of significant research attention. Additionally, we note that distribution systems can also ignite wildfires~\cite{russell2012}. Studying analogous formulations that aim to further mitigate wildfire risks and load shedding during PSPS events via infrastructure investments in distribution systems is an important direction for future research.

We also note that this paper uses the DC power flow approximation to formulate mixed-integer linear programming (MILP) problems that can be solved with commercial tools. Since the switching decisions from solutions to OTS problems that use a DC power flow approximation may be suboptimal or infeasible when evaluated using an accurate AC power flow model~\cite{potluri2012,coffrin2014switching,barrows2014}, our future work aims to extend the results in this paper by leveraging the wide range of power flow approximations and relaxations that have been developed in the last decade~\cite{molzahn_hiskens-fnt2019}.

This paper is organized as follows. Section~\ref{sec:psps} first introduces notation and formulates a multi-period extension of the wildfire-risk-minimizing OTS problem from~\cite{rhodes2020balancing}. Section~\ref{sec:investments} then proposes our models for infrastructure investments in batteries, solar PV generators, and infrastructure hardening. Sections~\ref{sec:data} and~\ref{sec:evaluation} next describe the test cases and procedures, respectively, that we use to evaluate our models, followed by the numerical results in themselves in Section~\ref{sec:results}. Section~\ref{sec:limitations} then discusses limitations of the proposed approach and needs for further research. Finally, Section~\ref{sec:conclusion} concludes the paper.

\section{Multi-Time-Period\\ Public Safety Power Shutoffs}\label{sec:psps}

In this section, we present a framework for optimized \textit{multi-time-period} PSPS events. Past work ~\cite{rhodes2020balancing} formulates optimized shutoffs for a single time period.
Here, we extend this existing work, and consider how varying demand throughout a day may affect the set of lines selected for de-energization.
In Section~\ref{sec:investments}, we investigate the placement of batteries to support line de-energization, which 
couples the time periods together.

Our framework models a single day of operation when the risk of wildfire ignition via power systems infrastructure is high.
Hence, system operators no longer operate the network to minimize generation cost, but rather to reduce the risk of wildfire ignition while minimizing load shedding.
As is consistent with United States Geological Survey (USGS) wildfire forecasts \cite{eidenshink2012united}, we assume that the risk associated with a single energized line is static over a day.  
Therefore, given a forecast for wildfire risk and multi-time-period demands, our framework selects lines to de-energize for an entire 24-hour period. 
Although switching lines multiple times per day would provide more flexibility for reducing load shedding, utilities must perform inspections of de-energized lines to ensure safe re-energizations~\cite{SCE_WildfirePlan}. The time required for these inspections precludes the use of intraday switching in our formulation.

We model power flow using the DC approximation, which neglects reactive power, line losses, and voltage magnitudes~\cite{stott2009}. 
Specifically, we use the $B\Theta$ representation of the DC power flow approximation.
Prior publications such as~\cite{potluri2012,coffrin2014switching,barrows2014} have highlighted discrepancies between OTS problems that use the DC approximation versus the AC power flow model. 
AC OTS problems are challenging mixed-integer nonlinear programming (MINLP) problems, and formulating scalable and computationally tractable solution methods are the focus of on-going research~\cite{potluri2012,coffrin2014switching,barrows2014,kocuk2017,bestuzheva2020}, with many approaches using power flow relaxations and approximations~\cite{molzahn_hiskens-fnt2019}. Therefore, we use the DC approximation for computational tractability in this investigative work, and will consider the AC power flow model in future extensions.

\subsection{Parameter and Variable Definitions}

For a given network, let $\mathcal{N}$ be the set of buses, $\mathcal{L}$ be the set of transmission lines, and $\mathcal{G}$ be the set of generators.
Let $\mathcal{T} = \{1, \ldots, T \}$ be the considered set of time indices over the period of a day, where $T$ is the final time period. We define a 100~MVA per unit (p.u.) base power. 
The following network parameters are provided for all lines $\ell \in \mathcal{L}$:
\begin{itemize}
    \item $b^\ell$, line susceptance in p.u.,
    \item $\overline{f}^\ell$, the power flow limit in p.u., 
    \item $r^\ell$, the wildfire risk incurred if line $\ell$ is energized as a unitless non-negative number,
    \item $n^{\ell, \text{fr}}$ and $n^{\ell, \text{to}}$,  \textit{to} and \textit{from} buses, respectively, where positive power flows from the \textit{from} bus to the \textit{to} bus,
    \item $\overline{\delta}^\ell$ and $\underline{\delta}^\ell$, upper and lower voltage angle difference limits in radians, respectively,
    \item $l^\ell$, line length in miles.
\end{itemize}
For all generators $i \in \mathcal{G}$, define the parameters:
\begin{itemize}
    \item $\overline{p}_g^i$ and $\underline{p}_g^i$, upper and lower power generation limits, respectively, in p.u.,
    \item $n^i$, bus at which generator $i$ is located.
\end{itemize}
For all buses $n \in \mathcal{N}$, define the parameters:
\begin{itemize}
    \item $p_{d,t}^n$, power demand at time period $t \in \mathcal{T}$ in p.u.,
    \item $\mathcal{G}^n$, the set of generators located at bus $n$,
    \item $\mathcal{L}^{n, {\text{to}}}$ and $\mathcal{L}^{n, {\text{fr}}}$, the subset of lines $\ell \in \mathcal{L}$ with bus $n$ as the designated \textit{to} bus, and bus $n$ as the designated \textit{from} bus, respectively.
\end{itemize}
The operation of the network during a multi-time-period PSPS event is characterized by the following set of variables using the B$\Theta$ representation of the DC power flow model:
\begin{itemize}
    \item $p_{g,t}^i$, power generated at unit $i \in \mathcal{G}$ at time period $t \in \mathcal{T}$ in p.u.,
    \item $\theta^n_t$, voltage angle at bus $n \in \mathcal{N}$ at time period $t \in \mathcal{T}$ in radians,
    \item $p_{ls,t}^n$, load shedding for buses $n \in N$ at time period $t \in \mathcal{T}$ in p.u.,
    \item $f^\ell_t$, power flowing from bus $n^{\ell, \text{fr}}$ to bus $n^{\ell, \text{to}}$ along line $\ell \in \mathcal{L}$ at time period $t \in \mathcal{T}$ in p.u.,
    \item $z^\ell \in \{0,1\}$, state of energization of line $\ell \in \mathcal{L}^{\text{switch}}$, where $\mathcal{L}^{\text{switch}} \subseteq \mathcal{L}$ is the subset of lines that can be de-energized. If $z^\ell=0$, then line $\ell$ is de-energized, and  if $z^\ell=1$, then line $\ell$ is energized. Note that the line's energization state is constant for all $t \in \mathcal{T}$.
\end{itemize}

\subsection{Operational and Physical Constraints}
We require that the generation at all units $i \in \mathcal{G}$ satisfy their lower and upper limits at each considered time index:
\begin{align} \label{const: gen limits}
    \underline{p}_{g}^i \leqslant p_{g,t}^i \leqslant \overline{p}_{g}^i, \qquad \forall i \in \mathcal{G}, \ \forall t \in \mathcal{T}.
\end{align}
The load shed at all buses and for all time periods must be positive\footnote{Some test cases model fixed generation as a negative load. We therefore only allow our formulation to shed loads that are positive, i.e., $d^n_t > 0$.} and cannot exceed the power demand at the bus:
\begin{align} \label{const: load shed limits}
    0 \leqslant p_{ls,t}^n \leqslant p_{d,t}^n, \qquad \forall n \in \mathcal{N}, \ \forall t \in \mathcal{T}.
\end{align}
The power flow along line $\ell \in \mathcal{L}$ must not exceed upper and lower line flow limits; however, if de-energized, then the power flow along line $\ell$ is zero:
\begin{align} \label{const: powr flow limits}
    -\overline{f}^\ell z^\ell \leqslant f^\ell_t \leqslant \overline{f}^\ell z^\ell, \qquad \forall \ell \in \mathcal{L}^{\text{switch}}, \ \forall t \in \mathcal{T}\\
    -\overline{f}^\ell \leqslant f^\ell_t \leqslant \overline{f}^\ell, \qquad \forall \ell \in \mathcal{L} \setminus \mathcal{L}^{\text{switch}}, \ \forall t \in \mathcal{T}.
\end{align}
For all lines, we require the voltage angle differences  to not exceed their lower and upper limits, unless the line is de-energized:
\begin{multline} \label{const: voltage angle}
    \underline{\delta}^\ell z^\ell + \underline{M}(1-z^\ell) \leqslant \theta^{n^{\ell, \text{fr}}}_t - \theta^{n^{\ell, \text{to}}}_t \leqslant \overline{\delta}^\ell z^\ell + \overline{M}(1-z^\ell)\\
    \qquad \qquad \forall \ell \in \mathcal{L}^{\text{switch}}, \ \forall t \in \mathcal{T},
\end{multline}
\begin{multline}
    \underline{\delta}^\ell  \leqslant \theta^{n^{\ell, \text{fr}}}_t - \theta^{n^{\ell, \text{to}}}_t \leqslant \overline{\delta}^\ell, \quad \forall \ell \in \mathcal{L}\setminus\mathcal{L}^{\text{switch}}, \ \forall t \in \mathcal{T},
\end{multline}
where $\overline{M}$ and $\underline{M}$ are big-M constants. For the numerical results in this paper, we compute these constants by simply summing the angle difference bounds across all lines. We note that more sophisticated approaches for computing these constants (e.g.,~\cite{fattahi2019,li2021}) could lead to faster solution times.

From the DC power flow approximation, the power flow on each line for each time period must abide by the following:
\begin{multline} \label{const: power flow}
    -b^\ell(\theta^{n^{\ell, \text{fr}}} - \theta^{n^{\ell, \text{to}}}) + |b^\ell|\underline{M}(1-z^\ell) \leqslant f_t^\ell\\
    \leqslant -b^\ell(\theta^{n^{\ell, \text{fr}}} - \theta^{n^{\ell, \text{to}}}) + |b^\ell|\overline{M}(1-z^\ell),\\
    \forall \ell \in \mathcal{L}^{\text{switch}}, \ \forall t \in \mathcal{T},
\end{multline}
\begin{multline}\label{const: power flow no big M}
    -b^\ell(\theta^{n^{\ell, \text{fr}}} - \theta^{n^{\ell, \text{to}}})  \leqslant f_t^\ell \leqslant -b^\ell(\theta^{n^{\ell, \text{fr}}} - \theta^{n^{\ell, \text{to}}}),\\
    \forall \ell \in \mathcal{L}\setminus\mathcal{L}^{\text{switch}}, \ \forall t \in \mathcal{T}.
\end{multline}
Note that if a line is energized (i.e., $z^{\ell}=1$),
then constraint \eqref{const: power flow} reduces to \eqref{const: power flow no big M}. 
Last, we require power balance at all buses for all time periods:
\begin{multline} \label{const: power balance}
    \sum_{\ell \in \mathcal{L}^{n, \text{fr}}} f^\ell_t - \sum_{\ell \in \mathcal{L}^{n, \text{to}}} f^\ell_t = \sum_{i \in \mathcal{G}^n} p_{g,t}^i - p_{d,t}^n + p_{ls,t}^n,\\
    \forall n \in \mathcal{N}, \ \forall t \in \mathcal{T}.
\end{multline}

\subsection{Objective Function} \label{sec: obj}
Our goal is to simultaneously minimize wildfire risk and load shedding, which are often competing objectives.
Let $D$ be the total demand in the network over all time periods, i.e.:
\begin{align}
    D = \sum_{t \in \mathcal{T}}\sum_{n \in \mathcal{N}} p_{d,t}^n, \nonumber
\end{align}
and let $R$ be the total wildfire risk the network poses if all lines $\ell \in \mathcal{L}$ are energized, i.e.:
\begin{align}
    R = \sum_{\ell \in \mathcal{L}} r^\ell. \nonumber
\end{align}
Now, let $\alpha \in [0,1]$ be a parameter that quantifies the priority of the user between these two competing objectives.
If $\alpha=1$, then the user is solely interested in reducing load shedding.
If $\alpha=0$, then the user is solely interested in reducing wildfire risk. 
For $\alpha \in (0,1)$, the user seeks a weighted balance of the two objectives.
For a given value of $\alpha$, let $C^\alpha(\cdot)$ be the objective that the user wishes to minimize, which is a function of the load shedding and line de-energizing variables:
\begin{align} \label{objective func}
    C^\alpha(z, p_{ls}) = \frac{\alpha}{D}\left(\sum_{t \in \mathcal{T}}\sum_{n \in \mathcal{N}} p_{ls,t}^n\right) + \frac{(1-\alpha)}{R}\left(\sum_{\ell \in \mathcal{L}} r^\ell z^\ell\right).
\end{align}
Observe that dividing the first and second terms in~\eqref{objective func} by the total demand $D$ and the total risk $R$ enables the interpretation of these terms as the fractions of load shed and wildfire risk remaining after the line switching operations.

\subsection{Multi-Time-Period PSPS Forumulation}
Now, we can formulate the multi-time-period PSPS optimization problem as:
\begin{equation}
\begin{aligned}
&\min\limits_{p_g, \theta, f, p_{ls}, z} \ \eqref{objective func} \ \ \text{s.t.} \ \eqref{const: gen limits} - \eqref{const: power balance}, \nonumber
\end{aligned}
\tag{MTP-PSPS}\label{MTP-PSPS}
\end{equation}
where our goal is to minimize a weighted combination of the total load shedding and wildfire risk over the traditional operational variables ($p_g$, $\theta$, and $f$), load shedding ($p_{ls}$), and the line switching decisions ($z$).

\section{Infrastructure Investments}\label{sec:investments}

The \eqref{MTP-PSPS} problem presented in the preceding section can help operators manage the trade-off between wildfire risk reduction and load shedding.
However, as the threat of wildfire ignition becomes more severe and the length of the wildfire season extends, it may be the case that no trade-off (i.e., no value of $\alpha$) provides acceptable system performance.
In this situation, communities may need to invest in new infrastructure that can either reduce wildfire risk directly (e.g., undergrounding lines) or support de-energizing additional lines via load shed reduction (e.g., installing grid-scale batteries).
California, for instance, is currently investing billions of dollars in such wildfire resilience infrastructure 
through state and federal funding~\cite{infrastructure_bill,ca_infrastructure_bill,pge_undergrounding_initiative,pge_undergrounding_initiative_nyt}.

In this section, we extend the \eqref{MTP-PSPS} problem formulation to consider the placement and operation of new infrastructure.
Although infrastructure placement decisions would ideally be made in a manner that accounts for many possible realizations of wildfire risk, in this investigative work, we make infrastructure placement decisions based on a representative wildfire risk realization due to the modeling and computational challenges that exist when jointly considering optimal switching and an infrastructure investment model. 
We present our development of a representative realization in Section \ref{sec:Investments_methodologies}, and discuss future research directions, including uncertainty modeling, in Section \ref{sec:limitations}.

We explore three types of investments: (1) grid-scale batteries, (2) solar PV, and (3) line hardening or line maintenance measures. 
We assume that a user of this investment framework is provided a budget for infrastructure improvements, and each investment has an associated cost. 
Therefore, investment decisions are based on the load shedding versus wildfire risk trade-off parameter $\alpha$, the total budget, the cost of the individual investments, and the representative wildfire risk realization.
We first discuss the three considered types of investments and then formulate the infrastructure investment problem.

\subsection{Grid-Scale Batteries}\label{sec:batteries}

First, we consider the placement of grid-scale batteries.
We allow any number of batteries to be placed at a subset of the buses $\mathcal{N}^{\text{batt}} \subseteq \mathcal{N}$.
We assume all batteries in the network have the following characteristics: 
\begin{itemize}
    \item $\overline{E}$ and $\underline{E}$, the maximum and minimum energy storage limits of the battery, respectively, in p.u.,
    \item $E^n_0$, the sum of the initial charges of the batteries at bus $n \in \mathcal{N}^{\text{batt}}$ in p.u.,
    \item $e \in (0, 1]$ charge efficiency, and $\frac{1}{e}$ is the discharge efficiency,
    \item $\overline{p}_c$ and $\underline{p}_c$, the maximum and minimum charge rate limits for a single battery, respectively, in p.u. per considered time interval,
    \item $\overline{p}_w$ and $\underline{p}_w$, are the maximum and minimum discharge rate limits for a single battery, respectively, in p.u. per considered time interval,
    \item $\phi^{\text{batt}}$, price of a single battery in millions of dollars.
\end{itemize}
For each bus $n \in \mathcal{N}^{\text{batt}}$, we introduce the following variables:
\begin{itemize}
    \item $x^n \in \mathbb{Z}$, number of batteries placed at bus $n$,
    \item $u^n_t \in \{0,1\}$, state of batteries located at bus $n$ at time $t \in \mathcal{T}$, where $u^n_t=1$ indicates that the batteries at bus $n$ are charging, and $u^n_t=0$ indicates discharging,
    \item $p_{c,t}^n$, charging rate at bus $n$ and at time $t \in \mathcal{T}$ in \mbox{p.u.} per considered time interval,
    \item $p_{w,t}^n$, discharging rate at bus $n$ and at time $t \in \mathcal{T}$ in p.u. per considered time interval.
\end{itemize}
We note that generalizations to consider multiple types of batteries with heterogeneous characteristics are straightforward.

Let $E_t^n(\cdot)$ be the total energy stored in all batteries placed at bus $n \in \mathcal{N}^{\text{batt}}$ at time $t \in \mathcal{T}$. The stored energy varies as the batteries charge and discharge:
\begin{multline}
    E_{t+1}^n(x,p_c,p_w) = x^nE^n_0 + \sum_{\tau = 1}^t e\cdot p_{c,\tau}^n -  \frac{1}{e}\cdot p_{w, \tau}^n\\
    \forall n \in \mathcal{N}^{\text{batt}}, \ \forall t \in \mathcal{T}. \nonumber
\end{multline}
The energy stored in the set of batteries at bus $n \in \mathcal{N}^{\text{batt}}$ must satisfy lower and upper storage limits:
\begin{align} \label{const: charge limits}
    x^n\underline{E} \leqslant E_{t+1}^n(x,p_c,p_w) \leqslant x^n\overline{E}, \qquad \forall n \in \mathcal{N}^{\text{batt}}, \ \forall t \in \mathcal{T}.
\end{align}
Recall that binary variable $u_t^n$ identifies the state of the set of batteries at bus $n$ as either charging ($u_t^n=1$) or discharging ($u_t^n=0$), which prevents a set of batteries from simultaneously charging and discharging.
This is enforced via lower rate limits:
\begin{equation} \label{const: charging rate}
\begin{aligned}
    p_{c,t}^n \geqslant u^n_t \underline{p}_c, \quad p_{w,t}^n \geqslant (1 - u^n_t) \underline{p}_w, \quad \forall n \in \mathcal{N}^{\text{batt}}, \ \forall t \in \mathcal{T}
\end{aligned}
\end{equation}
and upper rate limits:
\begin{alignat}{2}\nonumber
    & p_{c,t}^n \leqslant  \overline{p}_c M^{\text{batt}} u^n_t, \qquad && p_{w,t}^n \leqslant  \overline{p}_w M^{\text{batt}} (1 - u^n_t), \qquad \\ \nonumber
    & p_{c,t}^n \leqslant  \overline{p}_c x^n, &&  p_{w,t}^n \leqslant  \overline{p}_w x^n, \\ 
    & && \qquad\quad \forall n \in \mathcal{N}^{\text{batt}}, \ \forall t \in \mathcal{T},
\end{alignat}
where $M^{\text{batt}}$ is a big-M constant that is equal to the maximum number of batteries possibly placed.

\subsection{Solar PV}\label{sec:solar}

Second, we consider the placement of solar PV to support line de-energization.
We assume solar PV can be placed at a subset of the buses $\mathcal{N}^{\text{solar}} \subseteq \mathcal{N}$, and the user has access to the following parameters:
\begin{itemize}
    \item $S^n_t$, the maximum possible output of a unit of solar PV at bus $n \in \mathcal{N}^{\text{solar}}$ and time interval $t \in \mathcal{T}$ in p.u.,
    \item $\phi^{\text{solar}}$, price of 1 unit of solar PV in millions of dollars.
\end{itemize}
For all $n \in \mathcal{N}^{\text{solar}}$, we introduce the following variables:
\begin{itemize}
    \item $a^n \geqslant 0$, the number of 1-unit installations of solar PV at bus $n$,
    \item $p_{s,t}^n \geqslant 0$, the solar PV output at bus $n$ and time interval $t \in \mathcal{T}$.
\end{itemize}

We require that the solar PV output does not exceed its upper bound on the possible power production, which is a function of the amount of solar PV installed at that bus as well as the location and time of day:
\begin{align} \label{const: solar}
    p_{s,t}^n \leqslant S^n_t a^n, \qquad \forall n \in \mathcal{N}^{\text{solar}}, \ \forall t \in \mathcal{T}.
\end{align}
Observe that~\eqref{const: solar} permits solar production below the maximum possible power production, i.e., ``spilling'' solar.

\subsection{Line Hardening and Maintenance}\label{sec:line hardening}

The final type of investment we consider is line hardening or maintenance for wildfire risk reduction.
Let the subset $\mathcal{L}^{\text{harden}} \subseteq \mathcal{L}$ be the set of lines that are candidates for hardening/maintenance.
We assume users have access to the following parameters:
\begin{itemize}
    \item $\beta \in [0,1]$, the reduction in wildfire risk due to line hardening or maintenance measures,
    \item $\phi^{\text{harden}}$, price of line hardening or maintenance in millions of dollars per mile of line length.
\end{itemize}
For all $\ell \in \mathcal{L}^{\text{harden}}$, we introduce the following variable:
\begin{itemize}
    \item $y^\ell \in \{0,1\}$, a state of the line indicating whether the line has been hardened or maintained ($y^\ell=1$) or no measures are enacted on the line ($y^\ell=0$).
\end{itemize}

We assume that the entire length of line $\ell$ is hardened or has increased maintenance.
Hardening or performing increased maintenance on partial segments of lines may provide better outcomes by targeting improvements in specific areas; however, 
this does not change the fundamental characteristics of the problem and our formulation could be extended accordingly. 
We note that reference \cite{taylor2021wildfire} studies line upgrades on partial segments for wildfire risk mitigation, but does not include a power flow model.

Note that there is no benefit to simultaneously hardening and de-energizing a line since de-energizing a line reduces its risk to zero while hardening a line reduces the risk by $\beta$ at a cost of $\phi^{\text{harden}}$ per mile. We impose the following constraint so that the solver does not consider simultaneously de-energizing and hardening a line: 
\begin{align} \label{const: Line hardening}
    (1-z^\ell) + y^\ell \leqslant 1, \qquad \forall \ell \in \mathcal{L}^{\text{harden}} \cap \mathcal{L}^{\text{switch}}.
\end{align}


\subsection{Infrastructure Investment Formulation} \label{sec: invest form}

Now we can formulate the infrastructure investment problem.
We require power balance at all buses:
\begin{multline} \label{const: power balance investments}
    \sum_{\ell \in \mathcal{L}^{n, \text{fr}}} f^\ell_t - \sum_{\ell \in \mathcal{L}^{n, \text{to}}} f^\ell_t = -p_{d,t}^n + p_{ls,t}^n - p_{c,t}^n + p_{w,t}^n\\ 
    + p_{s,t}^n + \sum_{i \in \mathcal{G}^n} p_{g,t}^i, \quad \forall n \in \mathcal{N}, \ \forall t \in \mathcal{T},
\end{multline}
where, for notational simplicity, we assume that $p_{c,t}^n=0$ and $p_{w,t}^n=0$, $\forall t \in \mathcal{T}$ if $n \not\in \mathcal{N}^{\text{batt}}$. Similarly, we assume $p_{s,t}^n=0$, $\forall t \in \mathcal{T}$ if $n \not\in \mathcal{N}^{\text{solar}}$.

Let $B$ be the infrastructure investment budget in millions of dollars. We require the total investment to be within this budget, i.e.: 
\begin{align}\label{const: budget}
    \sum_{\ell \in \mathcal{L}^{\text{harden}}} \phi^{\text{harden}} l^\ell y^\ell + \sum_{n \in \mathcal{N}^{\text{batt}}} \phi^{\text{batt}} x^n + \sum_{n \in \mathcal{N}^{\text{solar}}} \phi^{\text{solar}}a^n \leqslant B.
\end{align}

We modify the objective function in \eqref{objective func} to include the reduction in wildfire risk via line hardening or maintenance.
Let $C^{\alpha}_{\text{invest}}(\cdot)$ be the modified objective function, which we define as:
\begin{align} \label{investment objective func}
    C^{\alpha}_{\text{invest}}(p_{ls}, z, y) = C^{\alpha}(p_{ls}, z) -  \frac{(1-\alpha)}{R} \sum_{j \in \mathcal{L}^{\text{harden}}} \beta r^j y^j.
\end{align}

We can now formulate the  multi-time-period PSPS infrastructure investment problem as:
\begin{equation}
\begin{aligned}
&\min\limits_{p_g, \theta, f, p_{ls}, z, x, u, p_c, p_w, a, p_s, y} \ \eqref{investment objective func}\\
&\qquad\qquad\text{s.t.} \ \eqref{const: gen limits} - \eqref{const: power balance},\; \eqref{const: charge limits} - \eqref{const: budget}, \nonumber
\end{aligned}
\tag{Invest-Opt}\label{Invest-Opt}
\end{equation}
where our goal is now to minimize a weighted combination of the total load shedding and wildfire risk taking into account the reduction of risk via line hardening or maintenance.
We optimize over the traditional operational variables ($p_g$, $\theta$, and $f$), load shedding ($p_{ls}$), line switching decisions ($z$), battery variables ($x$, $u$, $p_c$, and $p_w$), solar PV variables ($a$ and $p_s$), and line hardening/maintenance variables ($y$).

\section{Test Networks and Parameter Values}\label{sec:data}

In this section, we present the two networks that we use for numerical tests. We also assign values to the parameters described in Sections~\ref{sec:psps} and~\ref{sec:investments} for the \eqref{Invest-Opt} problem. 

\subsection{Networks}
\begin{figure}[t]
  \centering
  \includegraphics[width=.45\linewidth]{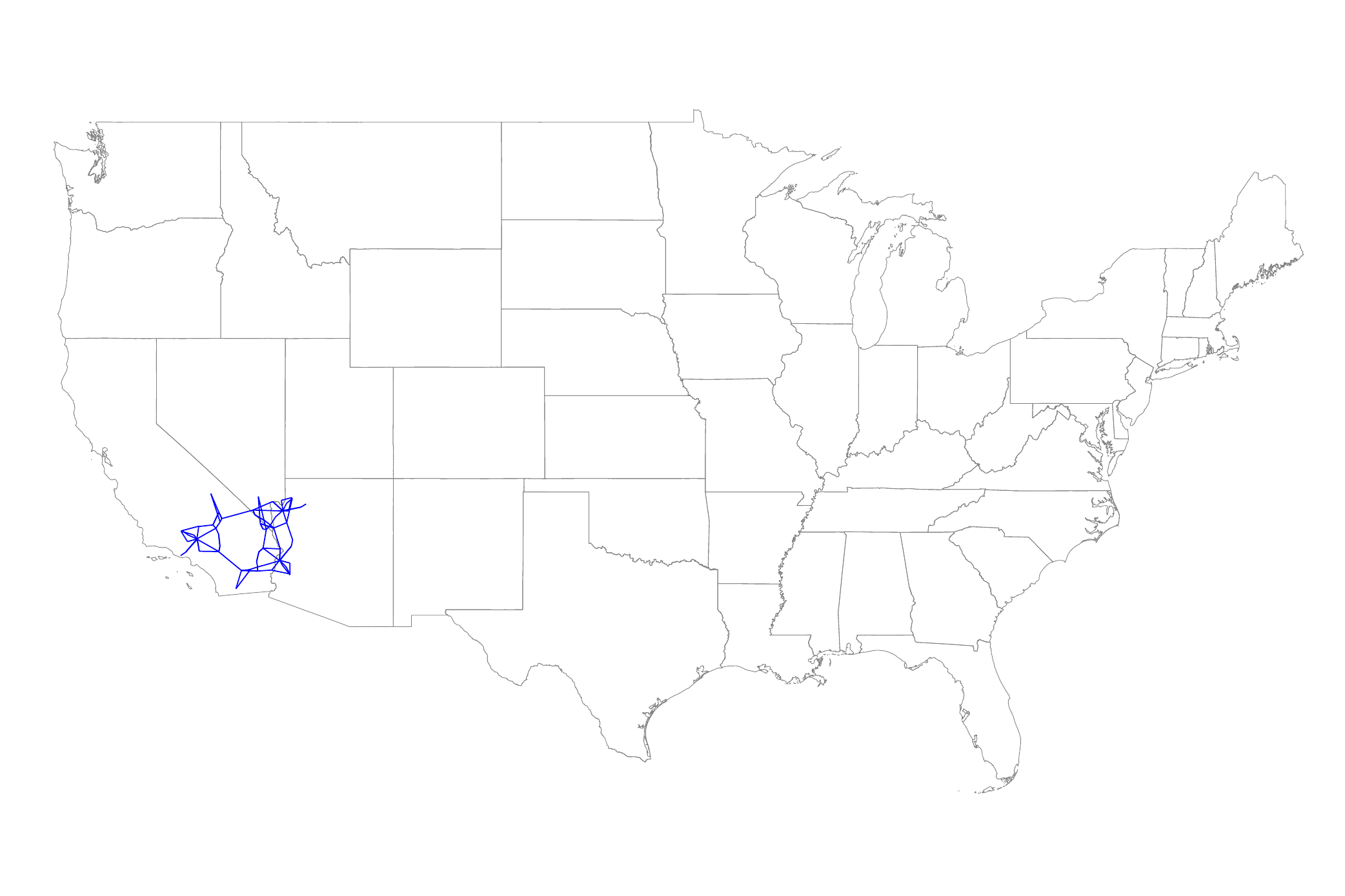}
  \includegraphics[width=.45\linewidth]{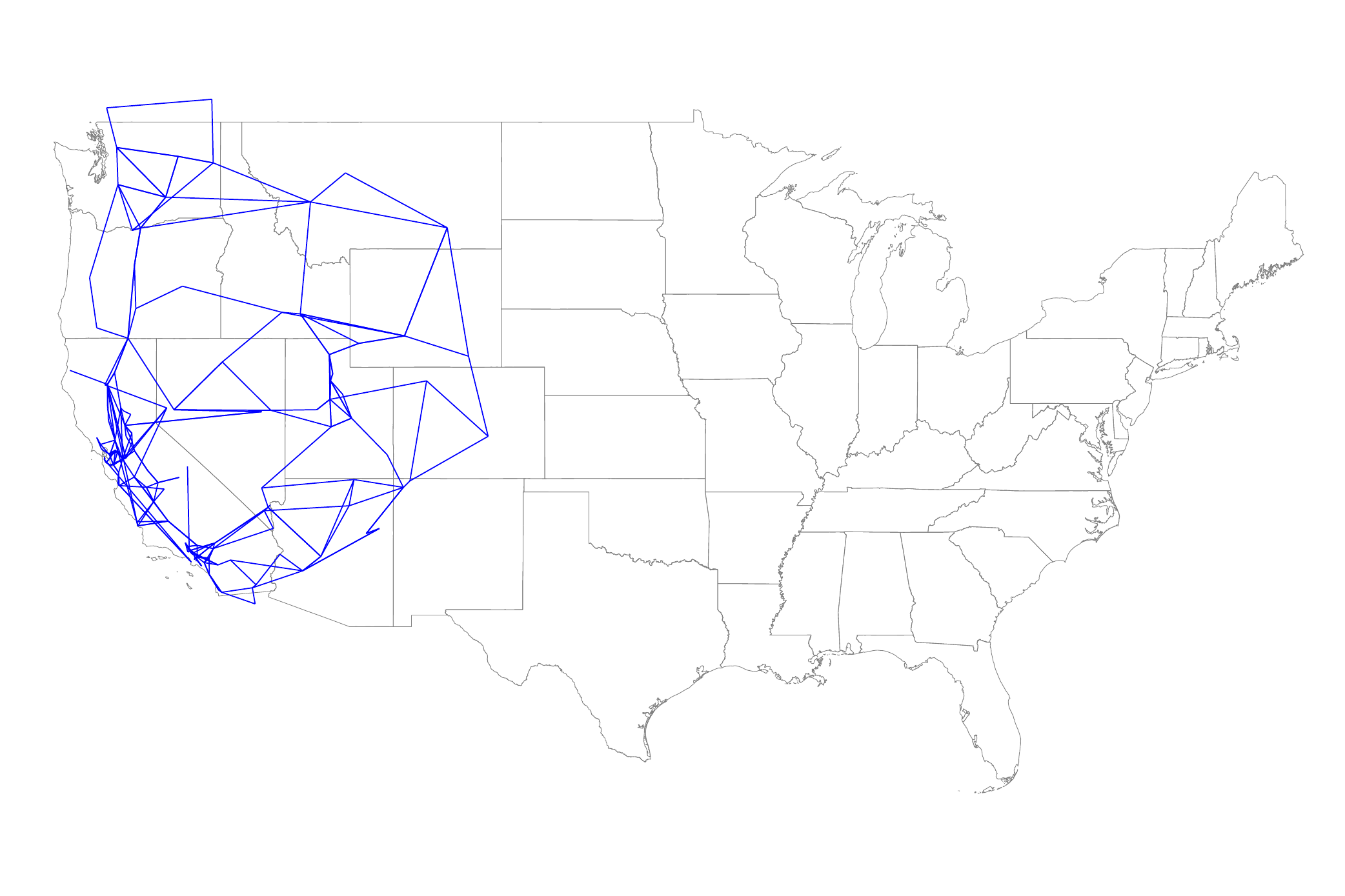}
  \caption{Locations of the RTS network (left) and the WECC network (right).}
  \label{figure:network locations}
\end{figure}

We demonstrate our algorithm using two synthetic transmission networks geolocated in parts of the western United States that intersect with historically high wildfire risk areas:
\begin{enumerate}
    \item RTS: 73-bus RTS-GMLC test case, Active Power Increase (API) version,
    \item WECC: 240-bus test case representing the Western Interconnect.
\end{enumerate}
The network topologies and electrical information associated with these test cases are adopted from~\cite{pglib-opf} based on data originating from~\cite{barrows2019ieee} and~\cite{price2011reduced}.
Figure \ref{figure:network locations} shows the locations of these networks within the continental United States.
Geographic locations of buses for the RTS network are available within the test case data; however, geographic data for the WECC network is not readily available. 
To geolocate most of the buses, we obtained partial geographic information from the WECC test case considered in~\cite{munoz2013}. 
For the remaining buses, we used the bus name and zone information provided in \cite{sarkar2013minding} to infer their locations.

Table \ref{table:components of networks} shows the number of buses, generators, and lines for the RTS and WECC networks.
For both considered networks, the lower limits of all generators are set to zero to guarantee solution feasibility, i.e., $\underline{p}_g^i = 0$, $\forall i \in \mathcal{G}$.
The WECC network has two DC lines located in southern California, which are modeled as pairs of negative and positive demands at the lines' terminals. To ensure solution feasibility, we assume these DC lines are not energized and thus remove the corresponding power injections.
Therefore, all demands in the network are non-negative.
We also assume a linear routing of transmission lines between the locations of their terminal buses.

\begin{table}[t]
  \centering
  \caption{Network Sizes}\label{table:components of networks}
\renewcommand{\arraystretch}{1.3}
\begin{tabular}{lc|c|c|}
\cline{3-4}
                                      &                 & RTS & WECC \\ \hline
\multicolumn{1}{|l|}{Number of buses} & $|\mathcal{N}|$ & 73  & 240  \\
\multicolumn{1}{|l|}{Number of generators} & $|\mathcal{G}|$ & 99  & 143  \\
\multicolumn{1}{|l|}{Number of lines} & $|\mathcal{L}|$ & 120 & 448  \\ \hline
\end{tabular}
\end{table}

\subsection{Wildfire Risk Values}

The wildfire ignition risk posed by an energized power line depends on a number of factors involving the environmental conditions around the line and the line's physical characteristics. Translating these factors into numeric risk values is challenging (see, e.g.,~\cite{waseem2021}) and requires detailed data that are not available for our test cases. 

As a proxy for more sophisticated calculations of the risk values, we use the daily forecasts of the Wind-enhanced Fire Potential Index (WFPI) produced by the USGS. WFPI is a unitless metric ranging from 0 to 150 that measures vegetation flammability while also accounting for wind speed, rain, temperature, etc. USGS reports that large fires are associated with the highest WFPI values~\cite{USGS_WFPI}. 

Historically, wildfire season in the western United States typically spans from late summer to early fall; however, recent wildfire seasons have been lengthening \cite{USDA_wildfire_season}.
Therefore, our analyses use data from June 1 to October 31, which we will refer to as the wildfire season.

For each day in wildfire season over the last three years (2019, 2020, and 2021), we assign a unitless wildfire risk value $r^\ell$ for each line $\ell \in \mathcal{L}$ in the considered networks. We calculate the risk value $r^\ell$ by integrating the WFPI forecast values along each line. 
This method inherently results in long lines having higher risk values, which is also a characteristic that utilities have correlated with high ignition risk~\cite{waseem2021}.

\subsection{Hourly Loads}
We consider time indices representing one-hour periods and select $T=24$ to model one day.
However, the RTS and WECC test cases in~\cite{pglib-opf} provide a single snapshot of nominal load demands. The multi-period optimization problem considered in this paper requires extending these test cases with time-varying load profiles. For this purpose, we modify the nominal loads according to the hourly, daily, and weekly scaling values reported in~\cite{rts96} to create hourly load profiles for each day during the wildfire season.

\subsection{Solar Parameters}
To model solar installations, each network is divided into various solar zones that group nearby buses in similar geographic regions.
A single bus near the center of each zone is selected to create a representative solar curve. A total of 3 zones are used for the RTS network and 14 zones for the WECC network. Using PVWatts from the National Renewable Energy Laboratory~\cite{pvwatts}, solar output from a fixed-tilt 1~kW solar panel is generated for each hour of each day in the wildfire season to determine the maximum solar output per unit of solar capacity installed at bus~$n$, $S^n_t$. Assuming linear scaling for output power, the total solar output for a node is then given by $S_t^n a^n$, where $a^n$ denotes the number of 1~kW solar panels installed at bus $n$.

\subsection{Battery Parameters}
Table~\ref{table:battery parameters} summarizes the battery parameters used in our numerical tests.
Batteries are modeled to have a capacity of 1.0 p.u. (100~MWh) in our problem formulation~\cite{sandia}. This is consistent with utility-scale lithium-ion battery installations~\cite{battery_landscape}. 
The charging efficiency, $e$, is 95\% for each battery installation~\cite{battefficiency}. Each battery is allowed to discharge completely and charge to 90\% in one time period with maximum charge and discharge rates of 0.95~p.u./hour~\cite{battefficiency}. 

\begin{table}[]
\renewcommand{\arraystretch}{1.3}
\caption{Battery Parameters}\label{table:battery parameters}
\centering
\begin{tabular}{r|c|c|}
\cline{2-3}
                                         & Parameter         & Value     \\ \hline
\multicolumn{1}{|r|}{minimum storage limit}  & $\underline{E}$   & 0 p.u.    \\
\multicolumn{1}{|r|}{maximum storage limit}  & $\overline{E}$    & 1.0 p.u.  \\
\multicolumn{1}{|r|}{charge efficiency}  & $e$               & 0.95 \\
\multicolumn{1}{|r|}{minimum charge rate}    & $\underline{p}_c$ & 0 p.u./hour    \\
\multicolumn{1}{|r|}{maximum charge rate}    & $\overline{p}_c$  & 0.95 p.u./hour  \\
\multicolumn{1}{|r|}{minimum discharge rate} & $\underline{p}_w$ & 0 p.u./hour    \\
\multicolumn{1}{|r|}{maximum discharge rate} & $\overline{p}_w$  & 0.95 p.u./hour  \\ \hline
\end{tabular}
\end{table}

\subsection{Risk Reduction Due to Line Hardening or Maintenance}

As discussed in Section \ref{sec:intro}, we consider three types of line hardening and maintenance investments: (1) undergrounding lines, (2) installing covered conductors, and (3) performing increased vegetation management. 
Recall from Section \ref{sec:line hardening} that $\beta \in [0,1]$ is a parameter that captures the reduction in wildfire risk due to the implementing one of the these three investments.
As shown by the $\beta$ values in Table~\ref{table:risks}, we assume that undergrounding a transmission line eliminates all wildfire ignition risk associated with that line, installing covered conductors reduces the risk by half, and performing increased vegetation management reduces the risk by a quarter based on estimates from the references provided in this table.

\begin{table}[h]
\centering
\renewcommand{\arraystretch}{1.2}
\caption{\label{table:risks} Risk Reduction via Line Hardening/Maintenance}
\begin{tabular}{r|c|c|}
\cline{2-3}
\multicolumn{1}{l|}{}                       & $\beta$ & References \\ \hline
\multicolumn{1}{|r|}{undergrounding}        & 1.0     &    \cite{cpuc}        \\
\multicolumn{1}{|r|}{covered conductors}    & 0.5     &   \cite{cpuc, WECC_webinar}         \\
\multicolumn{1}{|r|}{vegetation management} & 0.25    &   \cite{palaiologou2018using}      \\ \hline
\end{tabular}
\end{table}

\subsection{Investment Budget}\label{sec:budget}
We consider total budgets ranging from \$100 million to \$1 billion in increments of \$100 million. The magnitude of these budgets are based on typical investment estimates being discussed by policymakers~\cite{infrastructure_bill,ca_infrastructure_bill,pge_undergrounding_initiative,pge_undergrounding_initiative_nyt}. 

\subsection{Costs of Considered Infrastructure}

Table~\ref{table:costs} lists the costs of the considered investments as well as 
the references used to identify these cost values.
Note that the cost per mile to underground existing transmission lines widely varies.
The Edison Electric Institute reports that converting overhead transmission lines to underground ranges between \$1.3 and \$14.7 million per mile in 2022 dollars~\cite{hall2012out}.
We have chosen to use a cost of \$3 million/mile taking into account that~\cite{cpuc} reports the cost of undergrounding is approximately seven times the cost of installing covered conductors per mile along with the cost values reported in other sources~\cite{WI_underground}.

\begin{table}[h]
\centering
\renewcommand{\arraystretch}{1.2}
\caption{\label{table:costs}Cost of Infrastructure Investments}
\begin{threeparttable}[t]
\begin{tabular}{r|c|c|}
\cline{2-3}
                                                       & Cost                  & References \\ \hline
\multicolumn{1}{|r|}{battery\tnote{1}}               & \$20 million per battery  &     \cite{NREL_batt_cost}        \\
\multicolumn{1}{|r|}{solar PV\tnote{2}}                & \$940 per 1-kW-DC array         &  \cite{solar_cost}   \\
\multicolumn{1}{|r|}{undergrounding}        & \$3 million per mile      &    \cite{cpuc, WI_underground}         \\
\multicolumn{1}{|r|}{covered conductors}    & \$0.5 million per mile    &   \cite{cpuc, SCE_cc, MISO}          \\
\multicolumn{1}{|r|}{vegetation management\tnote{3}} & \$0.01 million per mile &    \cite{veg_cost}          \\ \hline
\end{tabular}
\begin{tablenotes}
     \item[1] 100 MWh lithium-ion grid-scale battery.
     \item[2] Fixed-tilt, utility-scale PV system.
     \item[3] Over a 20 year period.
   \end{tablenotes}
    \end{threeparttable}%
  \label{tab:addlabel}%
\end{table}

We have reported the cost of vegetation management for a 20-year period so that the cost is comparable to the lifetime of grid-scale lithium-ion batteries, roughly 15 years, and solar panels, roughly 25-30 years \cite{sandia, futurePV}.

\section{Evaluation and Benchmarking Methodologies}\label{sec:evaluation}

In this section, we describe our methodologies for selecting and evaluating the performance of infrastructure investment decisions.
We first determine the optimal types, quantities, and locations of new infrastructure based on a representative wildfire risk realization derived from 2019 and 2020 risk values and the optimization formulation described in Section~\ref{sec: invest form}. Next, we evaluate
these infrastructure investment decisions by simulating PSPS events during the 2021 wildfire season and analyze
the resulting wildfire risk reductions and load shedding.
In the following subsections, we describe the details of these methodologies.

\subsection{Infrastructure Investment Decisions}\label{sec:Investments_methodologies}

We examine eight infrastructure investment scenarios, each considering different combinations of infrastructure types, as outlined in Table~\ref{table:scenarios}.
For each scenario, we find the optimal placement via the \eqref{Invest-Opt} formulation for the ten budgets described in Section~\ref{sec:budget}.
For each budget, we evaluate $\alpha$ values ranging from $0.05$ to $0.95$ in increments of $0.05$. We do not present results for $\alpha=0$ or $\alpha=1.0$ because these values give unrealistic solutions (e.g., de-energize all lines). Thus, for each investment scenario, we assess 190 total cases.

\begin{table*}[h]
\centering
\renewcommand{\arraystretch}{1.2}
\caption{\label{table:scenarios} Investment Scenarios}
\begin{tabular}{|c|c|c|c|}
\hline
Scenario & Considered infrastructure types                           & Nonempty sets                                                                        & Empty sets \\ \hline
1    & batteries                                  & $\mathcal{N}^{\text{batt}}$                                                          & $\mathcal{N}^{\text{solar}}, \mathcal{L}^{\text{harden}}$\\
2    & solar PV                                   & $\mathcal{N}^{\text{solar}}$                                                         & $\mathcal{N}^{\text{batt}}, \mathcal{L}^{\text{harden}}$\\
3   & undergrounding                             & $\mathcal{L}^{\text{harden}}$                                                        & $\mathcal{N}^{\text{batt}}, \mathcal{N}^{\text{solar}}$\\
4    & covered conductors                         & $\mathcal{L}^{\text{harden}}$                                                        & $\mathcal{N}^{\text{batt}}, \mathcal{N}^{\text{solar}}$ \\
5    & vegetation management                      & $\mathcal{L}^{\text{harden}}$                                                        & $\mathcal{N}^{\text{batt}}, \mathcal{N}^{\text{solar}}$ \\
6   & batteries, solar PV, undergrounding        & $\mathcal{N}^{\text{batt}}, \mathcal{N}^{\text{solar}}, \mathcal{L}^{\text{harden}}$ & --  \\
7   & batteries, solar PV, covered conductors    & $\mathcal{N}^{\text{batt}}, \mathcal{N}^{\text{solar}}, \mathcal{L}^{\text{harden}}$ & --\\
8   & batteries, solar PV, vegetation management & $\mathcal{N}^{\text{batt}}, \mathcal{N}^{\text{solar}}, \mathcal{L}^{\text{harden}}$ & --\\ \hline
\end{tabular}
\vspace{-1em}
\end{table*}

Ideally, we would select infrastructure investments by jointly considering daily risks and load shedding for a full wildfire season or calendar year.
However, solving such an optimization problem that simultaneously considers $T = 24\cdot 365 = 8760$ hourly time periods is computationally intractable. 
Therefore, we aggregate the wildfire risk and load demand into a single representative worst-case date.
As future work, we intend to use stochastic optimization methods to better capture temporal variations and uncertainties in the wildfire risk, load demands, and solar power availability.

Since we are interested in placing infrastructure in a manner that minimizes wildfire risk and load shedding over a season, we develop a profile of wildfire risks to capture the historically riskiest values.
Recall that $r^{\ell}$ is the risk value associated with line $\ell \in \mathcal{L}$.
To assign a risk value to $r^\ell$ for the infrastructure investment problem, we average the top $10\%$ of all risks experienced by line $\ell$ over the 2019 and 2020 wildfire seasons. Since the infrastructure investment problem does not correspond to a particular day but rather a representative 24-hour period, the demands are assigned to be the nominal loads modulated by the hourly load profile of the peak demand day. Moreover, all batteries selected for placement in the network are initially fully charged in anticipation of a PSPS event.

\subsection{Performance Evaluation: 2021 Wildfire Season Simulation}
\label{sec:sequential_analysis}

After making infrastructure investment decisions for the scenarios outlined in the previous subsection based on risk data for the 2019 and 2020 wildfire seasons, we aim to analyze the success of these decisions through a sequential simulation of the 2021 wildfire season.
Denote the infrastructure investment decisions resulting from solving problem \eqref{Invest-Opt} as $x=\hat{x}$ (battery location and quantity), $a=\hat{a}$ (solar PV array location and quantity), and $y=\hat{y}$ (line hardening locations).

For each day in the 2021 wildfire season, we first determine if the wildfire threat is high enough to necessitate de-energizing lines via a threshold on the total risk during that day. 
Recall from Section~\ref{sec: obj} that $R$ is the total wildfire risk of the network if all lines remain energized. 
In our assessment methodology, operators are required to reduce the total risk of the network by making line de-energization decisions during any day for which $R\geqslant R_{\text{PSPS}}$, where $R_{\text{PSPS}}$ is a specified system-wide de-energization threshold. 
Conversely, if $R < R_{\text{PSPS}}$, then the risk the network poses is not great enough to require de-energizing lines.

We set $R_{\text{PSPS}}$ to represent the $75^{\mathrm{th}}$ percentile of the daily $R$ values from the 2019 and 2020 wildfire seasons.
Using this threshold to analyze the 2021 wildfire season, there are 14 and 28 days when the total wildfire risk exceeds this threshold for the RTS and WECC test cases, respectively.
This threshold was chosen because the resulting number of days when PSPS events occurred is similar to the number of events that have actually been enacted annually~\cite{sdge_psps_website,balaraman2021}.

For each day that meets or exceeds the $R_{\text{PSPS}}$ threshold, we use the wildfire risk values and demands associated with that particular date to make de-energization decisions. 
We also assume that any batteries in the network are able to fully charge before a PSPS event. However, if PSPS events occur on back-to-back days, we assume that the batteries may not have enough time to fully recharge after the first PSPS day. Thus, the initial states-of-charge for the batteries in such cases are set to the final states-of-charge from the proceeding day.

Line de-energization decisions are made to minimize the weighted sum of network load shedding and wildfire risk with investment decisions fixed. This method represents the behavior of a system operator who follows an optimal transmission switching strategy to balance wildfire risk and load shedding as in Section~\ref{sec:psps}, analogous to the proposal in~\cite{rhodes2020balancing}. 

Note that we modify the objective function in~\eqref{investment objective func} to include a term that incentivizes increased final states-of-charge for batteries to improve the battery flexibility on back-to-back days with PSPS events. Let the total available battery storage be:
\begin{align}
    E_{\text{total}} = \sum\limits_{n \in \mathcal{N}^{\text{batt}}} \hat{x}^n\overline{E}.
\end{align}
We also account for the reduction in wildfire risk due to line hardening investments.
Now, let the modified objective function be  $C^{\alpha}_{\text{seq}}(\cdot)$, which we define as:
\begin{multline} \label{seq objective func}
    C^{\alpha}_{\text{seq}}(p_{ls}, z, p_c, p_w) = \frac{\gamma}{E_{\text{total}}} \sum_{n \in \mathcal{N}^{\text{batt}}} E^n_{T+1}(\hat{x}, p_c, p_w)\\
    +  \frac{\alpha}{D}\left(\sum_{t \in \mathcal{T}}\sum_{n \in \mathcal{N}} p_{ls,t}^n\right) + \frac{(1-\alpha)}{R}\left(\sum_{\ell \in \mathcal{L}} r^\ell z^\ell \left( 1 - \beta \hat{y}^\ell\right) \right),
\end{multline}
where $\gamma \geqslant 0$ is a scaling term that we set to $\gamma = 0.01$. A small value of $\gamma$ incentivizes operation which stores energy for the next day to the extent that doing so does not overly restrict wildfire risk and load shedding performance for the current day. 
For notation simplicity, we assume $\hat{y} = 0$ for all $y \not\in \mathcal{L}^{\text{harden}}$.
With this modification, we formulate the de-energization problem with fixed infrastructure investments as:
\begin{equation}
\begin{aligned}
&\min\limits_{p_g, \theta, f, p_{ls}, z, x, u, p_c, p_w, a, p_s, y} \ \eqref{seq objective func}\\
&\qquad\qquad\text{s.t.} \ \eqref{const: gen limits} - \eqref{const: power balance},\; \eqref{const: charge limits} - \eqref{const: budget}, \nonumber\\
&\qquad\qquad\quad \ \; x=\hat{x}, \ a=\hat{a}, \ y=\hat{y}.
\end{aligned}
\tag{Seq-Opt}\label{Seq-Opt}
\end{equation}

\section{Numerical Results}\label{sec:results}

This section 
presents the results of the evaluation and benchmarking methodologies discussed in Section~\ref{sec:evaluation} applied to the two test cases described in Section~\ref{sec:data}.
Optimization problems were solved using Gurobi~9.1.0 to a 1\% MIP gap (except for a few instances that were unable to converge to this accuracy within a reasonable time period and which we identify accordingly). To implement the optimization formulations, we used Julia 1.6.1 with JuMP~0.22.2 along with the data input functionality of PowerModels.jl 0.19.1~\cite{coffrin2018}.

\subsection{RTS Results}

\begin{figure*}[tp]
  \centering
  \subfloat{\includegraphics[width=1\linewidth]{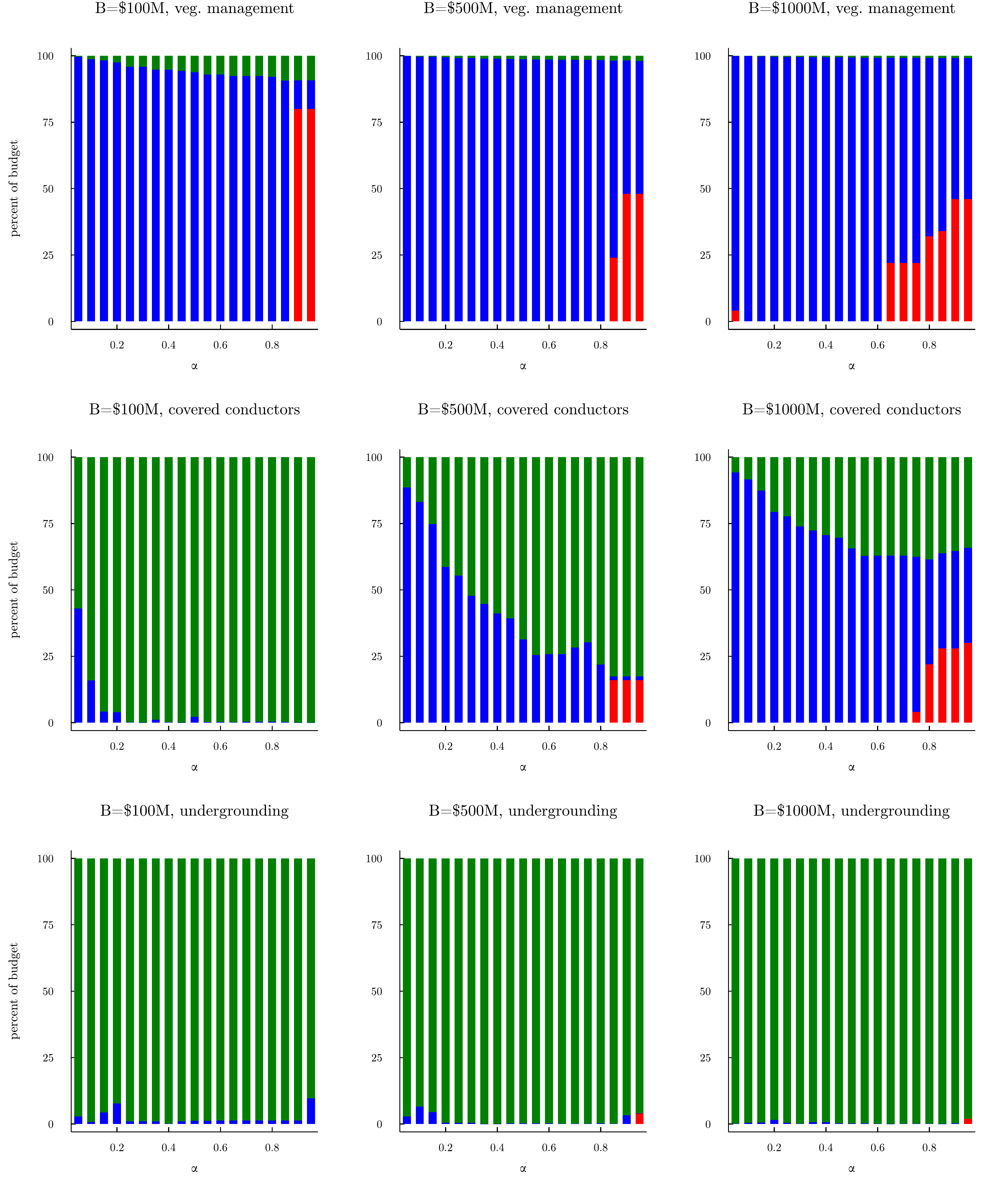}}
  \caption{Percent of budget spent on different investment types for the RTS network. Results are shown for three budgets ($\$100$M, $\$500$M and $\$1000$M) and three difference investment scenarios as described in Table \ref{table:scenarios} (Scenarios 6, 7, and 8). 
  The plots show the budget breakdown for various values of trade-off parameter $\alpha$ when the formulation is allowed to install batteries (red bars), solar PV (blue bars), and one of the three line hardening/maintenance options (green bars): increased vegetation management (top row), covered conductors (middle row), and undergrounding (bottom row).}
    \label{figure:RTS percent budget}
\end{figure*}

For the RTS network, we allow all lines to be de-energized or to be hardened/maintained, i.e., $\mathcal{L}^{\text{switch}}=\mathcal{L}^{\text{harden}}=\mathcal{L}$, and we allow solar PV and batteries to be placed at any bus, i.e., $\mathcal{N}^{\text{solar}}=\mathcal{N}^{\text{batt}}=\mathcal{N}$.
We first illustrate the performance of the \eqref{Invest-Opt} formulation for the RTS system. Figure~\ref{figure:RTS percent budget} shows the optimal budget allocation between grid-scale batteries (red bars), solar PV (blue bars), and hardened/maintained lines (green bars).
The plots in each column correspond to three different budget values: $B=\$100$M, $B=\$500$M, and $B=\$1000$M. 
Each row in this figure corresponds to a different investment scenario in Table \ref{table:scenarios}, which includes the possible installation of batteries, solar PV, and one of the three line hardening/maintenance options: increased vegetation management (Scenario~8, top row), covered conductors (Scenario~7, middle row), and undergrounding (Scenario~6, bottom row).
The horizontal axis of each plot is the considered $\alpha$ value and the vertical axis is the percent of the budget.
This figure shows that the optimal solution to \eqref{Invest-Opt} often decides to spend all or nearly all of the budget on undergrounding lines when given that option, showing that the benefits of completely eliminating wildfire ignition risk by undergrounding lines outweighs its high cost. 
A much smaller fraction of the budget is spent on intensive vegetation management; however, the cost per mile for vegetation management is much cheaper (0.333\%) than the cost per mile to underground lines, which likely explains why this accounts for a smaller fraction of the budget.
For the RTS network, batteries are almost exclusively installed for high values of $\alpha$ (prioritizing reduced load shedding). This may be because this case has more energized lines which enables more effective use of the batteries' charging and discharging capabilities.

\begin{figure*}[tp]
  \centering
  \subfloat{\includegraphics[width=1\linewidth]{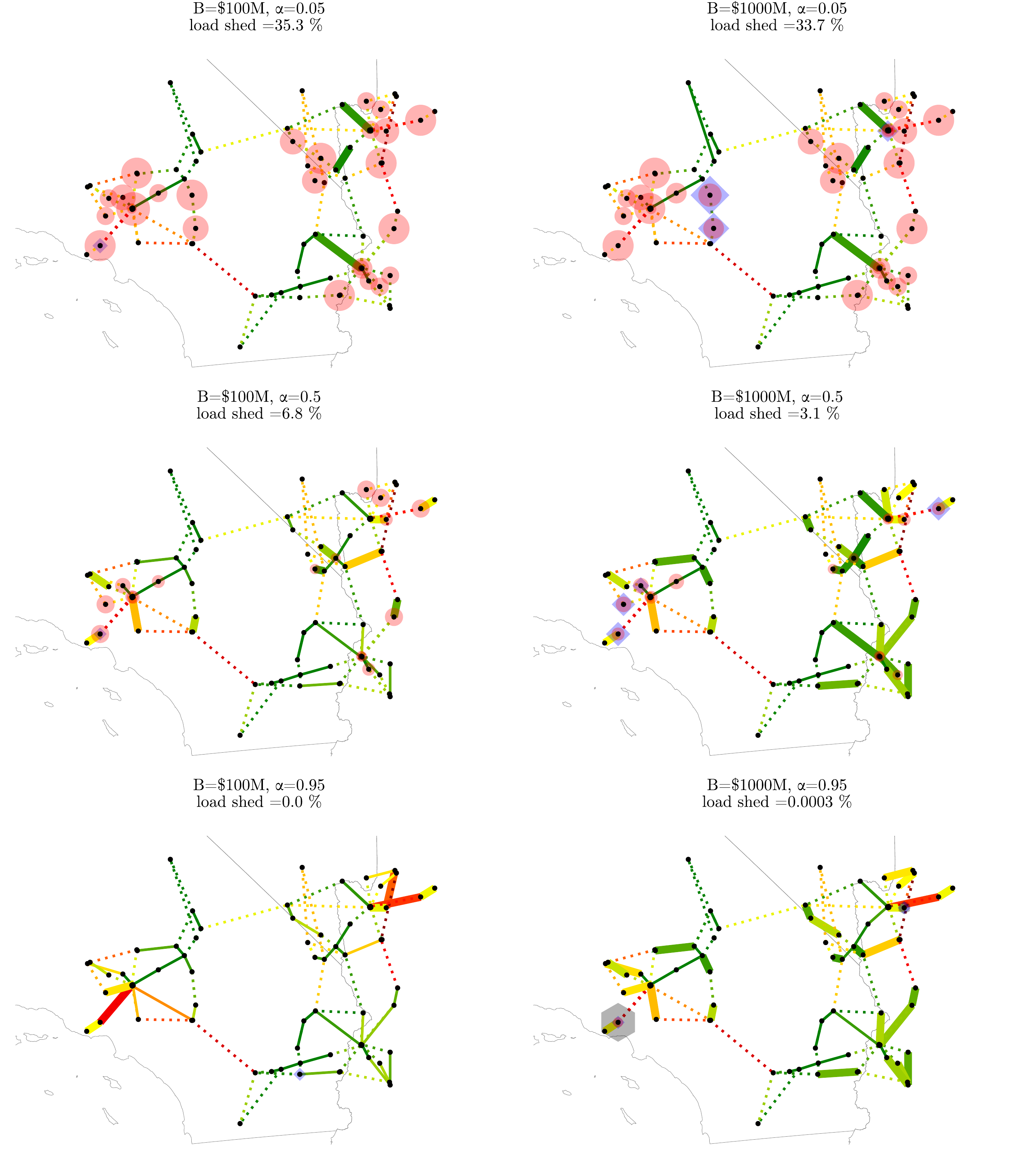}}
  \caption{Location of infrastructure investments for the RTS network if allowed to install batteries, solar PV, and covered conductors. The plots show results for two different budgets: $B=\$100$M (left column) and $B=\$1000$M (right column), as well as three different values of the trade-off parameter: $\alpha=0.05$ (top row), $\alpha=0.5$ (middle row), $\alpha=0.95$ (bottom row). 
  Red circles show the amount of load shedding at the associated bus. Larger circles indicate more load shedding. Grey hexagons and blue diamonds mark battery and solar PV installations, respectively. Again, larger symbols indicate more installations at that bus. The color of a transmission line illustrates the wildfire risk incurred if that line is energized. Dark red lines have the most risk, dark green lines have the least risk, and orange lines pose a medium risk. Lines that are dotted are selected to be de-energized, and thickened lines are selected to hardened via covered conductors.}
    \label{figure:RTS invstment locations}
\end{figure*}

Figure~\ref{figure:RTS invstment locations} displays the results of the \eqref{Invest-Opt} formulation on the network diagram with investment options for installing batteries, solar PV, and covered conductors with budgets $B=\$100$M and $B=\$1000$M and $\alpha$ values of $0.05$ (prioritize reductions in wildfire ignition risk), $0.5$ (equal weighting of priorities), and $0.95$ (prioritize reductions in load shedding). Lines are colored based on their risk value, where green corresponds to low risk and red corresponds to high risk.
Lines that are chosen to be hardened via covered conductor installation are thick, while unhardened lines are thin. Lines that are de-energized are dotted.
Red circles mark the load shedding at each bus, where larger circles indicate larger amounts of load shedding in absolute quantities (MW) as opposed to a percentage of the demand at the bus.
Blue diamonds mark solar PV installations and grey hexagons mark battery installations, where larger shapes indicate more installations at that bus.
Low values of $\alpha$ prioritize wildfire risk reduction, and therefore, we see much more load shed, regardless of budget, for $\alpha=0.05$.
Conversely, high values of $\alpha$ prioritize load shedding reduction, and we see small or no load shedding for $\alpha=0.95$. With many generators and a robust transmission network, we note that a large number of lines can be turned off in the RTS system to reduce wildfire risks with relatively limited load shedding.

\begin{figure*}[tp]
  \centering
  \subfloat{\includegraphics[width=1\linewidth]{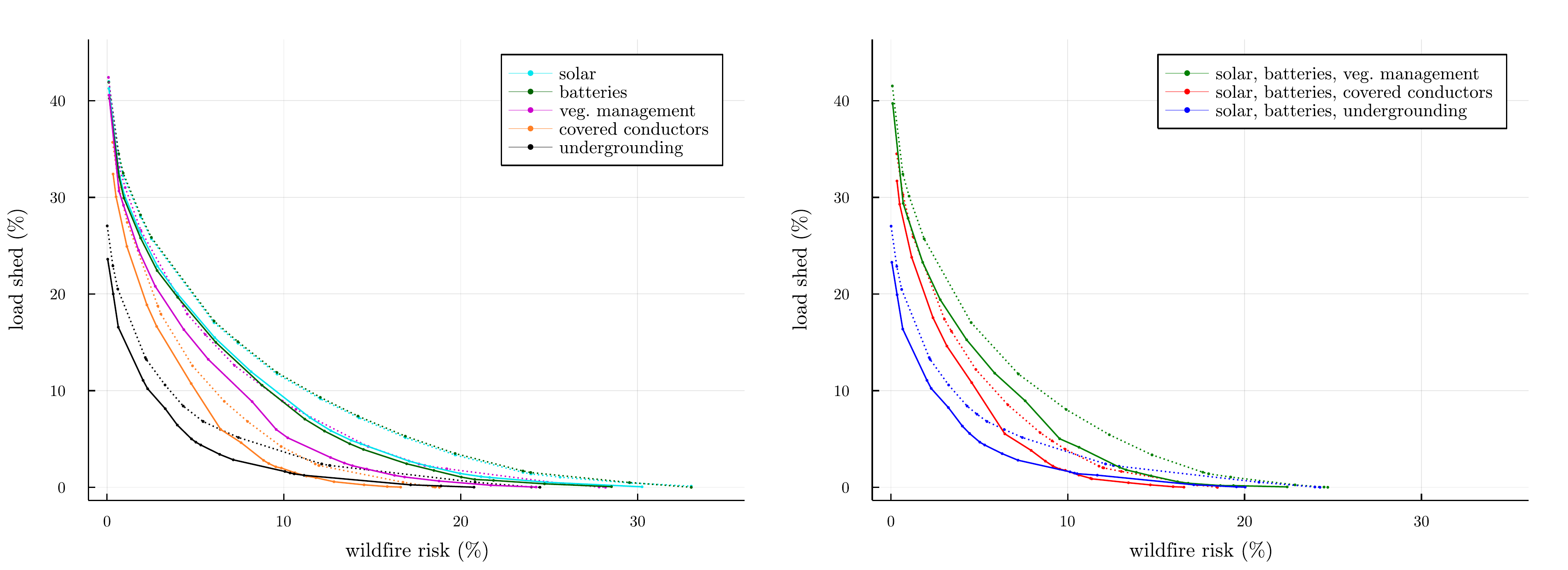}}
  \caption{Trade-off curves for single-investment strategies (left) and multi-investment strategies (right) for $B=\$500\text{M}$ for the RTS network. Dotted lines mark trade-off curves predicted from the placement optimization \eqref{Invest-Opt}. Load shedding and wildfire risk values are normalized by the total load and the total wildfire risk posed by the network in the worst-case manufactured risk and load profiles discussed in Section~\ref{sec:Investments_methodologies}. Solid curves result from the 2021 season-long simulation discussed in Section~\ref{sec:sequential_analysis}. Load shedding and wildfire risk values are normalized by the total load and the total wildfire risk posed by the network over the entire 2021 season of PSPS days.}
    \label{figure:RTS compare investments}
\end{figure*}

\begin{figure}[th!]
  \centering
  \subfloat{\includegraphics[width=1\linewidth]{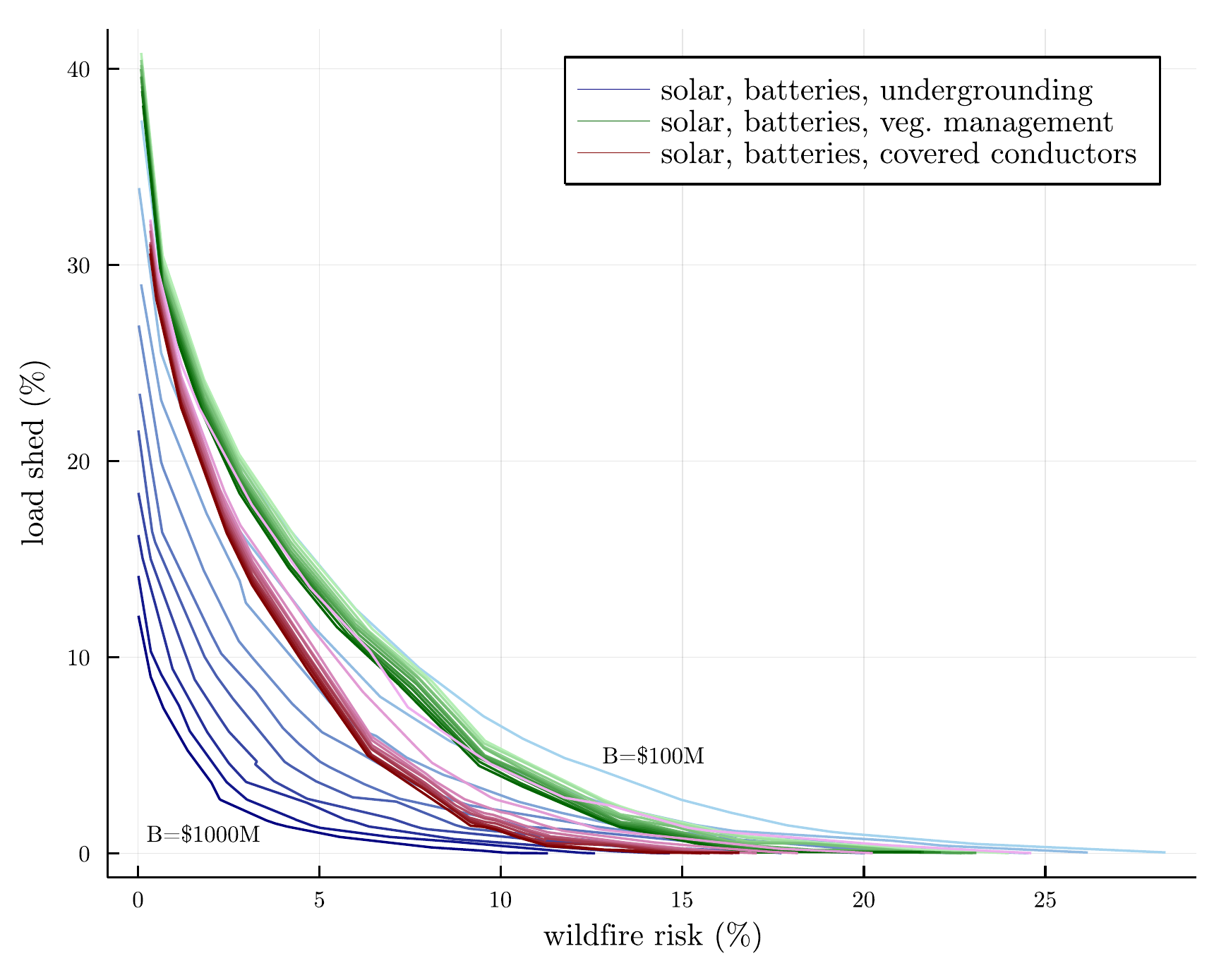}}
  \caption{Trade-off curves from season-long simulations of the RTS network for budgets ranging from $B=\$100\text{M}$ to $B=\$1000\text{M}$ in increments of $\$100\text{M}$. Results are shown when the formulation is allowed to install batteries, solar PV, and one of the three line hardening/maintenance options: increased vegetation management (green), covered conductors (red), and undergrounding (blue). Darker colored lines correspond to larger investment budgets, and lighter colored lines mark smaller budgets. For the undergrounding case (blue), the $B=\$100\text{M}$ and $B=\$1000\text{M}$ curves are annotated. Load shedding and wildfire risk values are normalized by the total load and the total wildfire risk posed by the network over the entire 2021 season of PSPS days.}
    \label{figure:RTS compare budgets}
\end{figure}

Figure~\ref{figure:RTS compare investments} illustrates trade-off curves of wildfire risk versus load shedding for the different investment scenarios in Table~\ref{table:scenarios} over values of $\alpha \in [0.05, 0.95]$ with $B=\$500$M. 
Lower values of $\alpha$ prioritize mitigation of wildfire risk at the top left of the plots. The curves slope down to the right as $\alpha$ increases to prioritize load shed reductions. 
As can be seen by comparing the right plot to the left plot, the trade-off curves for multi-investment strategies are largely dictated by the type of line hardening or maintenance selected, with the inclusion of solar PV and battery installations being relatively ineffective at shifting the curves.
Dotted lines mark the predicted trade-off curves resulting from the optimized placements (i.e., the solutions to \eqref{Invest-Opt}) using the representative wildfire risk profile (see Section \ref{sec:Investments_methodologies}), and solid lines mark the performance resulting from the 2021 season-long simulation (see Section~\ref{sec:sequential_analysis}).
Observe that the dotted curves are conservative estimates of the investments' performance in the 2021 wildfire season. This suggests that the wildfire risk profile used in the placement problem \eqref{Invest-Opt} overestimates the wildfire risks encountered during typical PSPS events throughout the season. This is expected given the intended bias in construction of these risk profiles towards high-risk days. We also note that relative performance of the investments are qualitatively similar for both the solid and dotted sets of curves.

Figure~\ref{figure:RTS compare budgets} shows the trade-off curves across all considered budgets, with higher budgets corresponding to darker lines on the plot. As the budget increases, the curves shift towards the origin which represents operations without load shed or wildfire risk. The largest improvement between budget values is seen in the case that includes undergrounding, which follows from the high expense and large wildfire risk reduction for this type of investment. Note that, at some values of $\alpha$ with a low budget, the case that includes undergrounding performs worse than the other cases. This may be because low budgets do not allow for many lines, or only short lines, to be undergrounded.

\begin{figure*}[t]
  \centering
  \subfloat{\includegraphics[width=1\linewidth]{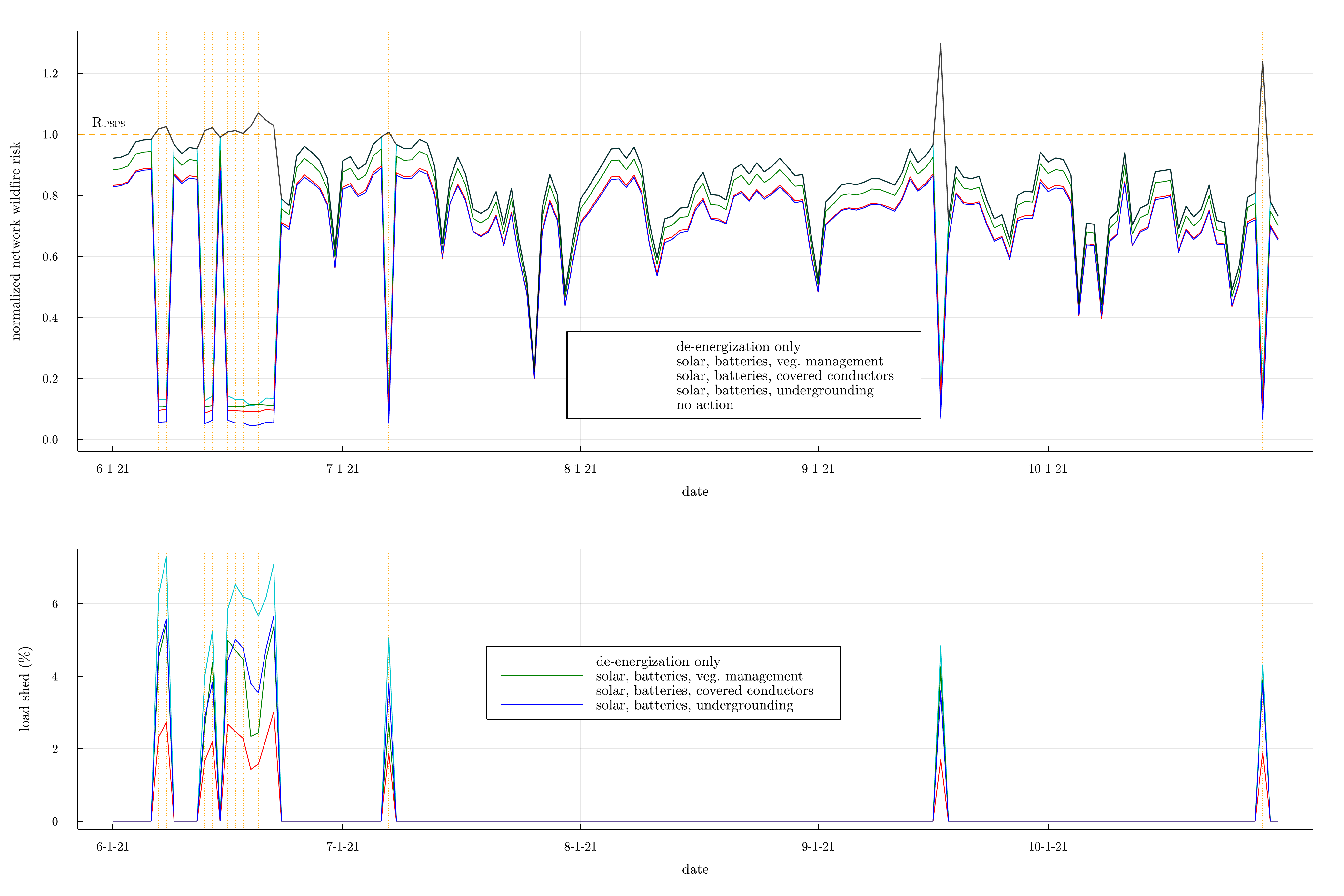}}
  \caption{Simulation of 2021 wildfire season for the RTS network given budget $B=\$500\text{M}$ and trade-off parameter $\alpha=0.5$. The top plot shows the wildfire risk for the 2021 wildfire season for various investment strategies. The horizontal dashed orange line marks the wildfire risk threshold, $R_{\text{PSPS}}$, which triggers a PSPS event. Vertical dotted orange lines mark days when the total wildfire risk $R$ is above the risk threshold $R_{\text{PSPS}}$. The black line marks the total risk posed by the network without any interventions (i.e., no de-energization or investments). Colored lines represent other possible strategies, including de-energization only (teal). All wildfire risk values are normalized by $R_{\text{PSPS}}$.
  The bottom plot shows the corresponding load shed as a percentage of the total load in the network on the associated day.}
   \label{figure:rts season performance}
\end{figure*}

Figure~\ref{figure:rts season performance} shows the wildfire risk and load shedding for the RTS network across the entire 2021 season with $B = \$500$M and $\alpha = 0.5$ under different investment strategies. The horizontal, dotted, orange line marks the threshold used to determine if a given day is a PSPS event. The PSPS events are determined based on the network wildfire risks before any investments are made, i.e., if investments lower the wildfire risk below the threshold on a given day, we still de-energize lines to have a consistent comparison. For the considered budget and $\alpha$, we note that while investments with undergrounding achieve the lowest wildfire ignition risks, investments with covered conductors result in the lowest amount of load shed during PSPS days. Recall that undergrounding eliminates a line's wildfire risk completely while covered conductors cut the risk in half at a fraction of the price.

\subsection{WECC Results}
\label{sec:wecc_results}

The increased size of the WECC network results in computational challenges when solving the \eqref{Invest-Opt} problem. To obtain results within reasonable computation times, we limit the sets of switchable lines, $\mathcal{L}^{\text{switch}}$, and buses where batteries could be placed, $\mathcal{N}^{\text{batt}}$. The 100 lines with the highest wildfire risk are allowed to be de-energized, and therefore, comprise the set $\mathcal{L}^{\text{switch}}$. As mentioned before, the problem formulation assigns higher risks to longer lines due to the integration method used in the risk assignment. Accordingly, the set of switchable lines is largely composed of the longest lines in the system. Based on observations of battery placements from solutions for the RTS system, we set $\mathcal{N}^{\text{batt}}$ to consist of the buses directly connected to a switchable line or any bus connected by a single line to those buses (i.e., all ``one-hop neighbors'' from the terminals of switchable lines). This results in 174 possible battery locations. 
We do not limit the placement of solar PV or line hardening/maintaining, i.e., $\mathcal{N}^{\text{solar}}=\mathcal{N}$ and $\mathcal{L}^{\text{harden}}=\mathcal{L}$.
The WECC network is otherwise evaluated using the same methodologies as the RTS network.
Note that four cases failed to achieve a 1\% MIP gap within a reasonable time period.
All four cases are investment scenario 8 (battery, solar PV, and intensive vegetation management) with budgets $B=\$\{600, 700, 800, 1000\}$M and $\alpha=\{0.90, 0.90, 0.85, 0.95\}$ with MIP gaps of \{1.51\%, 1.70\%, 1.23\%, 1.48\%\}, respectively.

\begin{table*}[t]
\centering
\renewcommand{\arraystretch}{1.7}
\caption{\label{table:num vars and const} Number of variables and constraints in placement optimization}
\begin{threeparttable}[t]
\begin{tabular}{ll|l|c|c|c|}
\cline{3-6}
                                                             &            &   Parameter expression                                                                                                                                                                                                                   & RTS & WECC & WECC limited\tnote{2}  \\ \hline
\multicolumn{1}{|c|}{\multirow{2}{1.5cm}{Number of variables}}   & Continuous & $|\mathcal{G}||\mathcal{T}| + 2|\mathcal{N}||\mathcal{T}| + |\mathcal{L}||\mathcal{T}|+ 2|\mathcal{N}^\text{batt}||\mathcal{T}| + |\mathcal{N}^\text{solar}|(1+|\mathcal{T}|)$                                                                                                           & 14089 & 42168 & 39000 \\
\multicolumn{1}{|c|}{}                                       & Integer    & $|\mathcal{L}^{\text{switch}}| + |\mathcal{N}^\text{batt}|(1+|\mathcal{T}|) + |\mathcal{L}^\text{harden}|$                                                                                                                           & 2056 & 6896 & 4898 \\ \hline
\multicolumn{1}{|l|}{\multirow{3}{1.5cm}{Number of constraints}} & Bound\tnote{1}     & $2|\mathcal{G}||\mathcal{T}| + 2|\mathcal{N}||\mathcal{T}| + 2|\mathcal{L}\setminus\mathcal{L}^{\text{switch}}||\mathcal{T}|$                                                                                                                               & 8256 & 16272 & 32976 \\
\multicolumn{1}{|l|}{}                                       & Inequality & $4|\mathcal{L}^{\text{switch}}||\mathcal{T}| + 4|\mathcal{L}\setminus\mathcal{L}^{\text{switch}}||\mathcal{T}| + 8|\mathcal{N}^{\text{batt}}||\mathcal{T}| + |\mathcal{N}^{\text{solar}}||\mathcal{T}| + |\mathcal{L}^{\text{harden}} \cap \mathcal{L}^{\text{switch}}|$ & 27408 & 95296 & 82276 \\
\multicolumn{1}{|l|}{}                                       & Equality   & $|\mathcal{N}||\mathcal{T}|$                                                                                                                                                                                                      & 1752 & 5760 & 5760 \\ \hline
\end{tabular}
\begin{tablenotes}
     \item[1] Lower and upper bounds on individual variables.
     \item[2] Assuming $|\mathcal{L}^{\text{switch}}|=100$, $|\mathcal{N}^{\text{batt}}|=174$, $|\mathcal{L}^{\text{harden}}|=448$, and $|\mathcal{N}^{\text{solar}}|=240$ (see discussion in Section~\ref{sec:wecc_results}). 
   \end{tablenotes}
    \end{threeparttable}
\end{table*}

\begin{figure*}[tp]
  \centering
  \subfloat{\includegraphics[width=1\linewidth]{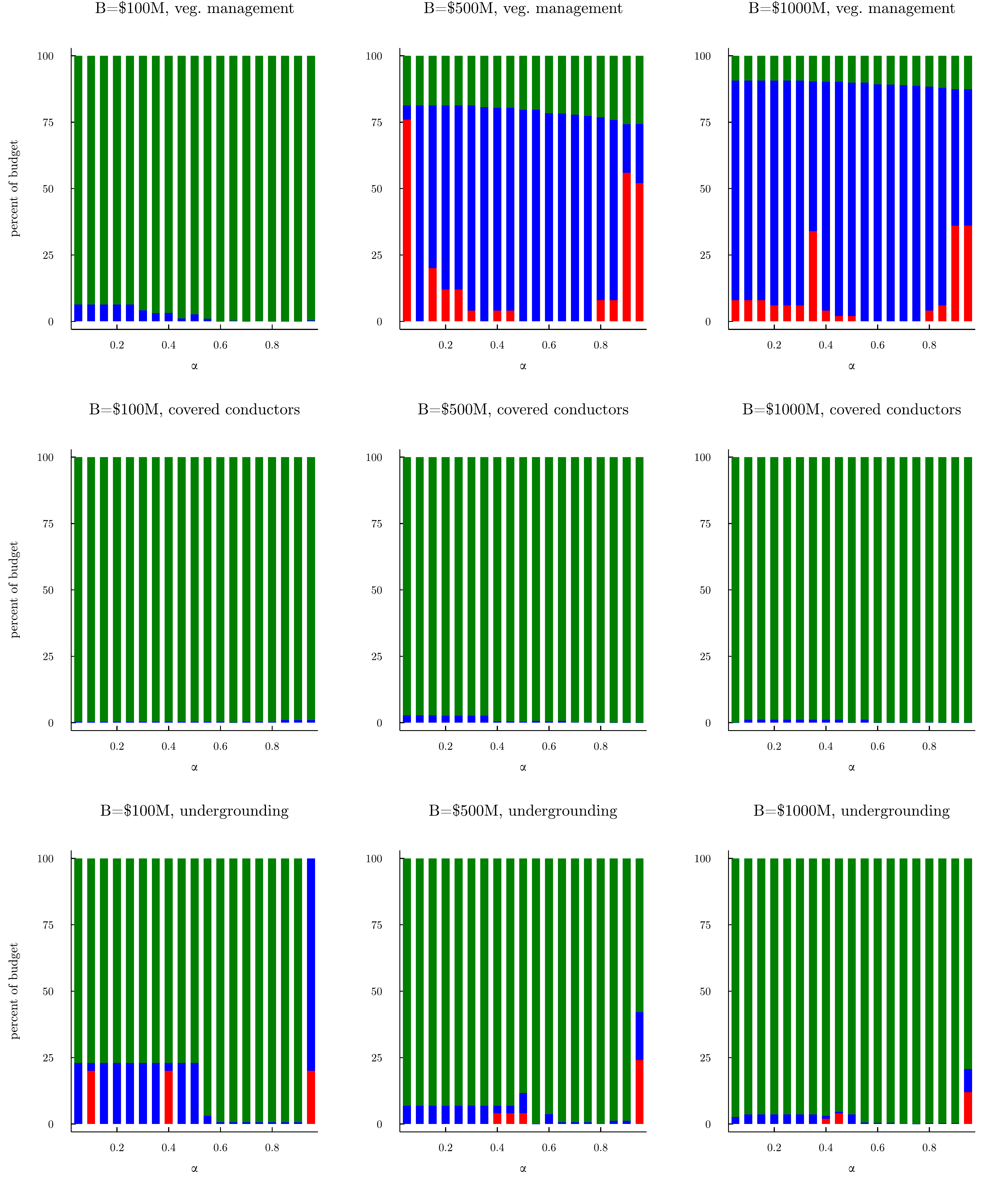}}
  \caption{Percent of budget spent on different investment types for the WECC network.
  Results are shown for three budgets ($\$100$M, $\$500$M and $\$1000$M) and three difference investment scenarios as described in Table \ref{table:scenarios} (Scenarios 6, 7, and 8). 
  The plots show the budget breakdown for various values of trade-off parameter $\alpha$ when the formulation is allowed to install batteries (red bars), solar PV (blue bars), and one of the three line hardening/maintenance options (green bars): increased vegetation management (top row), covered conductors (middle row), and undergrounding (bottom row).}
    \label{figure:percent budget wecc}
\end{figure*}

\begin{figure*}[tp]
  \centering
  \subfloat{\includegraphics[width=1\linewidth]{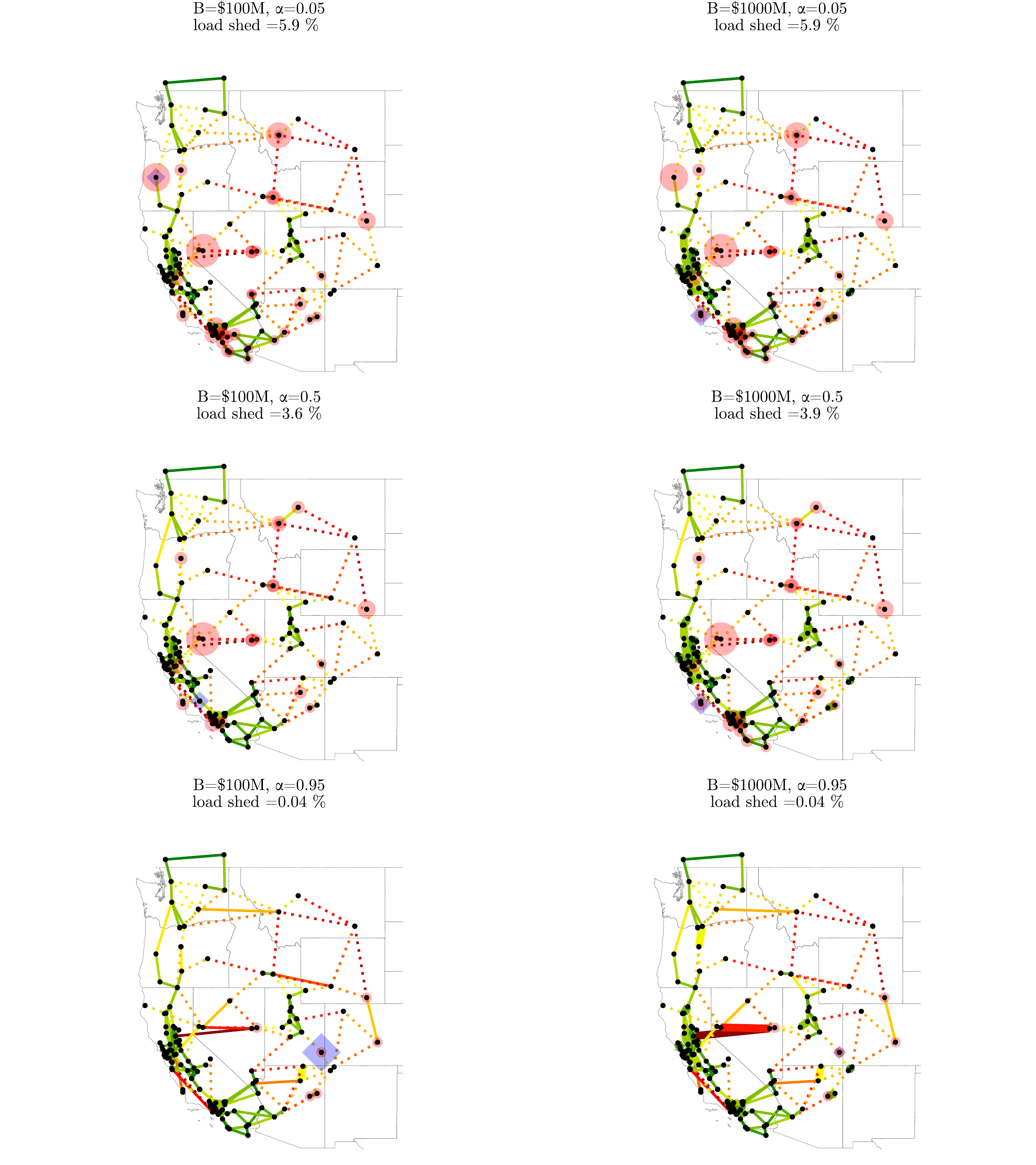}}
  \caption{Location of infrastructure investments for the WECC network if allowed to install batteries, solar PV, and covered conductors. The plots show results for two different budgets: $B=\$100$M (left column) and $B=\$1000$M (right column), as well as three different values of the trade-off parameter: $\alpha=0.05$ (top row), $\alpha=0.5$ (middle row), $\alpha=0.95$ (bottom row). 
  Red circles show the amount of load shedding at the associated bus. Larger circles indicate more load shedding. Grey hexagons and blue diamonds mark battery and solar PV installations, respectively. Again, larger symbols indicate more installations at that bus. The color of a transmission line illustrates the wildfire risk incurred if that line is energized. Dark red lines have the most risk, dark green lines have the least risk, and orange lines pose a medium risk. Lines that are dotted are selected to be de-energized, and thickened lines are selected to hardened via covered conductors.}
    \label{figure:wecc investment locations}
\end{figure*}

Figure~\ref{figure:percent budget wecc} shows the optimal budget allocation between grid-scale batteries, solar PV, and hardened/maintained lines across three combinations of investments and three budget values.
Figure~\ref{figure:wecc investment locations} shows the placements considering investments in batteries, solar PV, and covered conductors with budgets $B=\$100$M and $B=\$1000$M, and $\alpha$ values of $0.05$, $0.5$, and $0.95$. Note that lower values of $\alpha$ have fewer hardened lines and more de-energized lines. Since lower $\alpha$ values prioritize mitigating wildfire risk, this implies a preference to shut off a line entirely rather than spend money to partially reduce the risk. This highlights the importance of considering optimal switching when working on investment problems as the investment outcomes are predicated on the possibility of line de-energization. We also note that many scenarios result in a significant number of de-energized lines due to the simplified nature of this representative system. While very large numbers of de-energized lines would not be realistic in a practical setting, this test case is still useful for illustrating the performance of the proposed \eqref{Invest-Opt} formulation. Furthermore, the results suggest regions to target for line de-energizations and infrastructure investments when using a more realistic and detailed dataset.

\begin{figure*}[tp]
  \centering
  \subfloat{\includegraphics[width=1\linewidth]{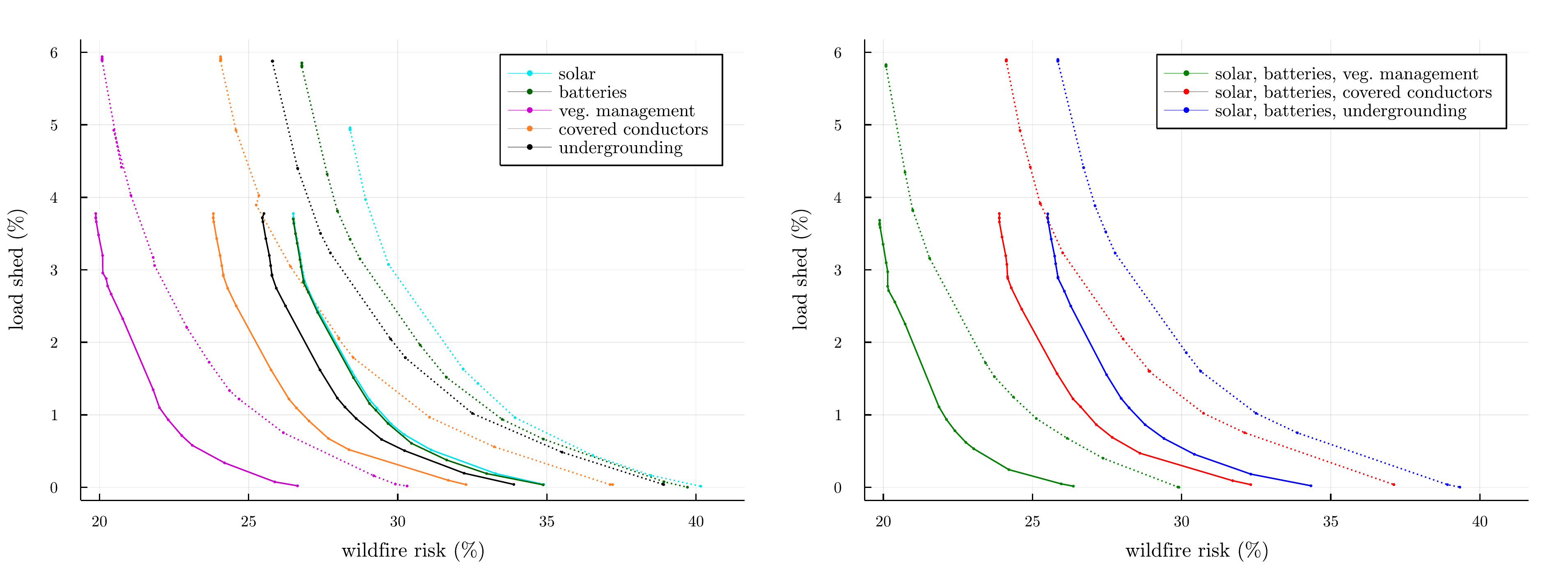}}
  \caption{Trade-off curves for single-investment strategies (left) and multi-investment strategies (right) for the WECC network with $B=\$500\text{M}$. Dotted lines mark trade-off curves predicted from the placement optimization \eqref{Invest-Opt}. Load shedding and wildfire risk values are normalized by the total load and the total wildfire risk posed by the network in the worst-case manufactured risk and load profiles discussed in Section~\ref{sec:Investments_methodologies}. Solid curves result from the 2021 season-long simulation discussed in Section~\ref{sec:sequential_analysis}. Load shedding and wildfire risk values are normalized by the total load and the total wildfire risk posed by the network over the entire 2021 season of PSPS days.}
    \label{figure:WECC compare investments}
\end{figure*}

\begin{figure}[]
  \centering
  \subfloat{\includegraphics[width=1\linewidth]{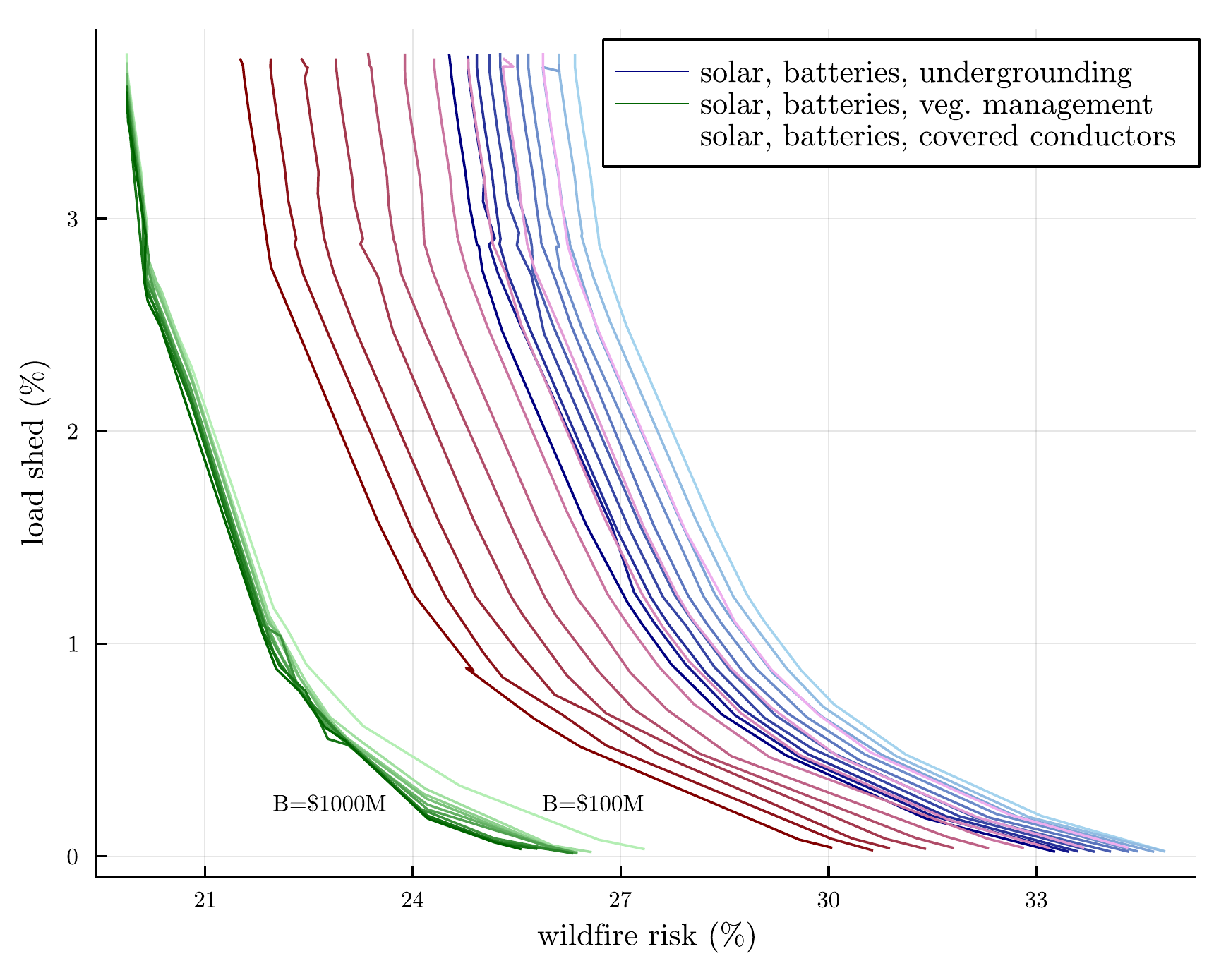}}
  \caption{Trade-off curves from season-long simulations for the WECC network with budgets ranging from $B=\$100\text{M}$ to $B=\$1000\text{M}$ in increments of $\$100\text{M}$. Results are shown when the formulation is allowed to install batteries, solar PV, and one of the three line hardening/maintenance options: increased vegetation management (green), covered conductors (red), and undergrounding (blue). Darker colored lines correspond to larger investment budgets, and lighter colored lines mark smaller budgets. For the vegetation management case (green), the $B=\$100\text{M}$ and $B=\$1000\text{M}$ curves are annotated. Load shedding and wildfire risk values are normalized by the total load and the total wildfire risk posed by the network over the entire 2021 season of PSPS days.}
    \label{figure:WECC compare budgets}
\end{figure}

Figure~\ref{figure:WECC compare investments} illustrates the trade-off curves for different infrastructure options, again with solid lines indicating the performance over the entire wildfire season and dashed lines indicating expected performance from the optimal investments with $B=\$500$M and $\alpha \in [0.05, 0.95]$. Once again, we note that the method of line hardening or maintenance largely determines the solution's performance when multiple investments are considered. However, unlike the RTS network, here undergrounding performs worse than both vegetation management and covered conductors. This is likely caused by the large size of the WECC network, containing over 27,500 miles of modeled transmission lines, and the limited budget available. With a budget of \$500 million and costs of \$3 million per mile for undergrounding and \$0.01 million per mile for vegetation management, at most 166.7 miles of lines can be undergrounded, while 50,000 miles can undergo increased vegetation management. This accounts for 0.6\% and 181.4\% of the total miles of lines in the WECC network, respectively. Being able to invest in vegetation management for large portions of this network could account for the increase seen in the performance of this line maintenance option. Similar trends are seen across all budgets as shown in Figure~\ref{figure:WECC compare budgets}.

\begin{figure*}[tp]
  \centering
  \subfloat{\includegraphics[width=1\linewidth]{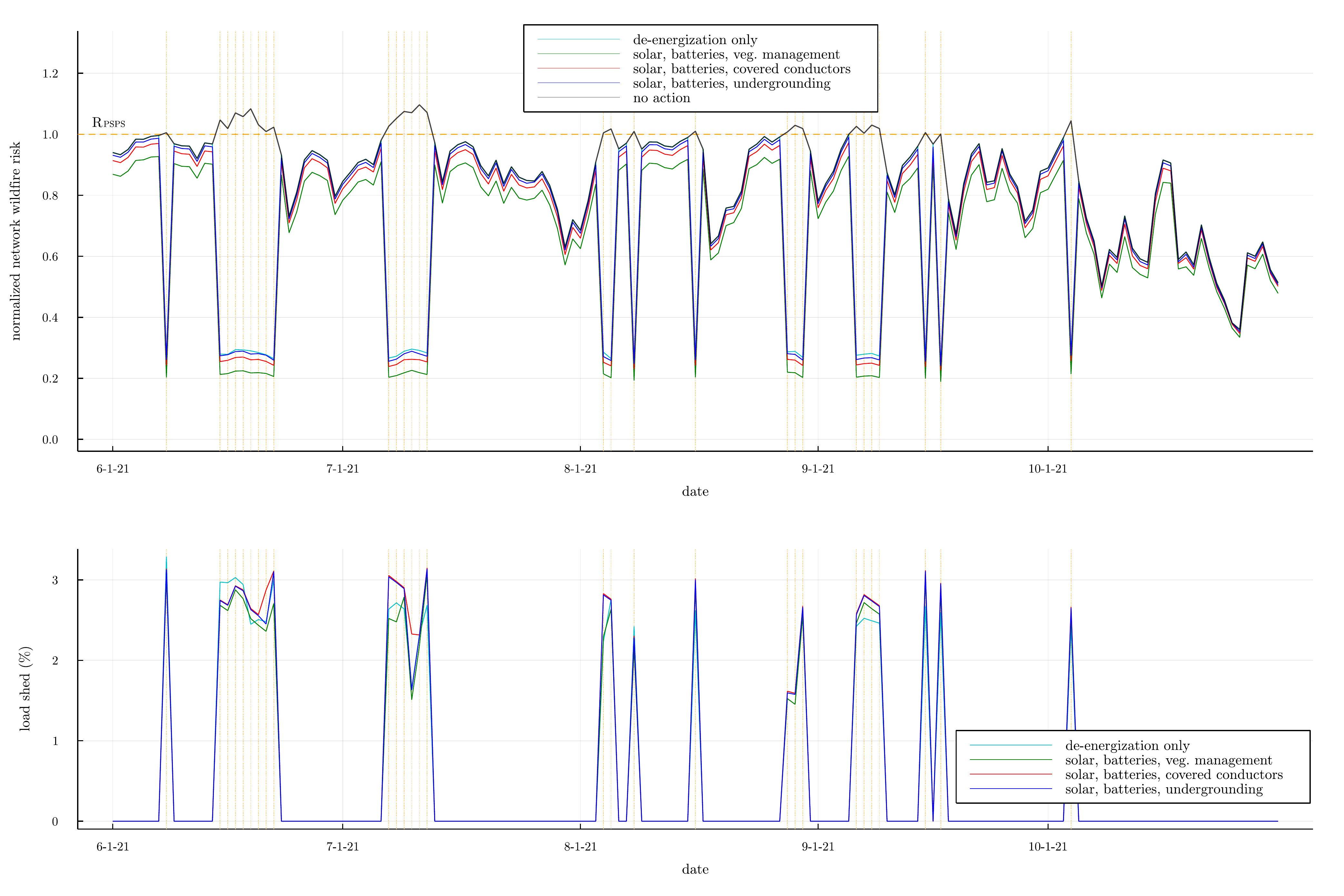}}
  \caption{Simulation of 2021 wildfire season for the WECC network given a budget $B=\$500\text{M}$ and trade-off parameter $\alpha=0.5$. The top plot shows the wildfire risk for the 2021 wildfire season with various investment strategies. The horizontal dashed orange line marks the wildfire risk threshold, $R_{\text{PSPS}}$, which triggers a PSPS event. Vertical dotted orange lines mark days when the total wildfire risk $R$ is above the risk threshold $R_{\text{PSPS}}$. 
  The black line marks the total risk posed by the network without any interventions (i.e., no de-energization or investments). Colored lines represent other possible strategies, including de-energization only (teal). All wildfire risk values are normalized by $R_{\text{PSPS}}$.
  The bottom plot shows the corresponding load shed as a percentage of the total load in the network on the associated day.}
    \label{figure:WECC season performance}
\end{figure*}

Finally, Figure~\ref{figure:WECC season performance} shows the performance of the WECC network across the 2021 wildfire season considering investment strategies including batteries and solar PV with either undergrounding, covered conductors, or vegetation management. Again, PSPS events are marked with vertical dotted lines when the network risk, without considering reductions from investments, is above the threshold $R_\text{PSPS}$ marked as a dashed, horizontal, orange line. As seen previously, investment strategies with vegetation management perform best by consistently having the lowest wildfire risk compared to other strategies and an average or lower amount of load shedding.

\subsection{Computational Complexity}
\label{sec:computational_complexity}
Similar to expansion planning problems in many settings, solving our infrastructure investment problem \eqref{Invest-Opt} is computationally challenging. Table~\ref{table:num vars and const} shows the number of variables and constraints associated with \eqref{Invest-Opt}. While the number of variables and constraints scale linearly with the size of the system, the problem is still difficult to solve for large cases, thus necessitating restrictions to the set of switchable lines and candidate locations for batteries as described above for the WECC test case.

Solver times are difficult to interpret precisely because our computations were conducted using a cluster with shared resources; however, in general, we found that scenarios with small to moderate values of $\alpha$, between 0.05 and 0.75, were completed in approximately one hour. Larger values of $\alpha$, between 0.8 and 0.95, took much longer, usually tens of hours, and many outlier cases took several days. As will next be discussed in Section~\ref{sec:limitations}, these timing results indicate that future work is needed to improve computational tractability.

\section{Limitations and Research Needs}\label{sec:limitations}

The approach to infrastructure investment that we present in this paper has advantages in terms of 
modeling flexibility and data availability. The numerical results give significant insights and show that accounting for line de-energization is crucial for appropriately choosing infrastructure investments. However, the investigatory work in this paper employs a number of simplifications that motivate the following future research directions:
\begin{itemize}
    \item \textbf{AC power flow models}: The  DC power flow model yields a tractable MILP formulation for the infrastructure investment problem. However, DC power flow inaccuracies may lead to unacceptable errors or even infeasibility when evaluated using an AC power flow model~\cite{potluri2012,coffrin2014switching,barrows2014}. Since directly using an AC power flow approximation results in a computationally challenging MINLP formulation, future work should study the trade-offs in solution quality and computational speed associated with various power flow approximations and relaxations~\cite{molzahn_hiskens-fnt2019}. 
    \item \textbf{Uncertainty models}: This paper formulates a deterministic problem with known values for load demands, solar generation, and wildfire risks. However, these aspects of the problem are all actually uncertain, motivating the use of stochastic optimization techniques. For instance, an extension to this work could jointly consider multiple scenarios for wildfire risks rather than a single aggregated scenario. Chance constrained and robust optimization techniques could also be used to account for uncertainty in load demands and solar generation.  
    \item \textbf{Longer time horizons}: The 24-hour horizon for the multi-period problem in this paper is based on both computational considerations and the fact that the wildfire risk values are constant for each day. However, this limited horizon necessitates the use of heuristics to model realistic battery behavior at the end of the horizon (see Section~\ref{sec:sequential_analysis}) and precludes modeling future periods using short-term wildfire risk forecasts. Extensions to longer time horizons would thus improve modeling realism.
    \item \textbf{More realistic generator models}: The conventional generator models in this paper are only constrained by bounds on their outputs at each time period. With more realistic modeling, future work could study how the constraints on ramp rates, minimum up and down times, reserve requirements, etc. that are found in unit commitment problems affect the load shedding versus wildfire risk trade-offs and investment decisions.
    \item \textbf{Controlled islanding}: The line de-energization model in this paper does not prevent the network from separating into multiple islands. Islanding can bring a range of operational challenges (e.g., stability considerations) that are not modeled in our formulation. Ongoing work is investigating the use of anti-islanding constraints that enable more flexible control of islanding behavior.
    \item \textbf{Contingencies}: The optimization formulation studied in this paper does not consider the impacts from failures of generators or lines. Future work could extend the formulation to model $N-1$ security constraints corresponding to the failure of any one individual component. Such extensions would raise questions such as how to appropriately penalize post-contingency load shedding.
    \item \textbf{Generalization to multiple planning periods}: The infrastructure investment problem considered here installs components and hardens infrastructure once and then evaluates the system's performance. Future work could study the impacts of allowing multiple rounds of investments along with progressively worsening wildfire conditions, changing infrastructure costs, and varying budgets over multiple years.
    \item \textbf{Multiple uses for infrastructure}: Investments in batteries, solar PV, and infrastructure hardening provide benefits during both wildfire conditions and normal conditions. This raises questions regarding valuations of these investments in problems that mitigate wildfire ignition risks. For more realistic results, future work should explore these questions by assessing the value of these investments during various operating conditions. 
    Moreover, future work should study how well investments intended for other objectives (e.g., generation cost minimization) perform in a wildfire setting (and vice-versa). 
    \item \textbf{Distribution system extensions}: Along with transmission systems, distribution systems also pose wildfire ignition risks. Analogs of the work in this paper could be developed for distribution systems via network reconfiguration~\cite{mishra2017}, the installation of microgrids with customer-scale battery storage and solar PV generation~\cite{moreno2022}, and the deployment of mobile distributed generators~\cite{taheri2021}.
    \item \textbf{Fairness considerations}: The problem formulation in this paper focuses on the system-wide performance metrics of total wildfire risk reduction and total load shedding. However, the impacts of wildfires and load shedding are often localized to particular parts of the system, which raises concerns related to the fairness of line de-energization outcomes and infrastructure investment decisions. Our ongoing work is studying how different operational and investment decisions affect individual loads with a particular focus on trade-offs among various metrics of system-wide performance and fairness.
\end{itemize}

Computational tractability is a key challenge associated with all of these research directions. Even without considering the extensions above, the large-scale MILP formulation in this paper can be challenging to solve, as discussed in Section~\ref{sec:computational_complexity}. Each extension could impose further computational burdens. This motivates the development of more effective heuristics for reducing the number of candidate locations for infrastructure improvements and possible line de-energization. To prioritize lines for de-energization, it may be possible to adapt heuristics from the optimal transmission switching literature (e.g.,~\cite{liu2012,fuller2012,ruiz2012}). However, since prior heuristics focus on network congestion effects that are less important in the wildfire risk setting, it is not clear whether they would be useful. Thus, further computational improvements are needed.

\section{Conclusions}\label{sec:conclusion}
With climate change amplifying the frequency and severity of wildfire conditions, power system operators de-energize transmission lines to mitigate acute wildfire ignition risks during PSPS events. There is a widely recognized need for new infrastructure investments to reduce the load shedding associated with these events. Local supplies of power from grid-scale batteries and solar PV installations can decrease the severity of outages when lines are de-energized. Hardening and maintaining transmission lines via undergrounding, installing covered conductors, and intensely managing vegetation reduces the wildfire ignition risks associated with the lines that remain energized. 

To consider spatially and temporally varying wildfire risks, load demands, and solar PV outputs, engineers require new computational tools to optimally locate and size infrastructure investments. Accordingly, this paper proposed a MILP formulation that models a multi-time-period representation of a power system during severe wildfire conditions. Our formulation locates batteries, solar PV, and line hardening/maintenance to reduce both wildfire ignition risks and load shedding.

Evaluations using two test cases with actual wildfire risk data from 2021 demonstrate the capabilities of the proposed formulation. Our results show the formulation's ability to choose among the investment options to obtain effective solutions tailored to the available budget, wildfire risks, and prioritization of load shedding versus wildfire ignition risk. Another key observation regards the high value of line hardening and maintenance activities, as these often account for significant fractions of the available budget. The results also illustrate the importance of jointly considering line de-energization and line hardening investments, as the solver de-energizes certain high-risk lines to completely eliminate their associated risks while allocating more of the budget to hardening moderate-risk lines. 

Finally, the paper describes several directions for extending this investigatory work to increase its accuracy and practical applicability. Many of these extensions would require improved computational tractability, which we emphasize as a key challenge for future research.

\section*{Acknowledgment}
The authors would like to thank Qingyu Xu, Yinong Sun, and Benjamin Hobbs for sharing geographic data regarding the 240-bus WECC test case.

This research was supported in part through research cyberinfrastructure resources and services provided by the Partnership for an Advanced Computing Environment (PACE) at the Georgia Institute of Technology.



\bibliographystyle{IEEEtran}
\IEEEtriggeratref{67}
\bibliography{IEEEabrv,refs.bib}
%

\end{document}